\documentclass[a4paper,11pt]{article}
\usepackage{jheppub} 

\newtheorem{theorem}{Theorem}

\def\U{\mathrm{U}}

\newtheorem{prop}{Conjecture}

\def\beq#1\eeq{\begin{align}#1\end{align}}

\title{Singularity, Sasaki-Einstein manifold, Log del Pezzo surface and $\mathcal{N}=1$ AdS/CFT correspondence: Part I}

\author[a,b]{Dan Xie}
\author[c,d,e]{Shing-Tung Yau}
\affiliation[a]{Yau Mathematics Science Center, Tsinghua University, Beijing, 10084, China}
\affiliation[b]{Department of Mathematics, Tsinghua University, Beijing, 10084, China}

\affiliation[c]{Department of Mathematics, Harvard University, Cambridge, MA 02138, USA}
\affiliation[d]{Center of Mathematical Sciences and Applications, Harvard University, Cambridge, 02138, USA}
\affiliation[e]{Jefferson Physical Laboratory, Harvard University, Cambridge, MA 02138, USA}

\abstract{A five dimensional Sasaki-Einstein (SE) manifold provides a AdS/CFT pair for four dimensional $\mathcal{N}=1$ SCFT, and those pairs are very useful in studying field theory and AdS/CFT correspondence. The space of known SE manifolds is increased significantly in the last decade, and we initiated the study of various field theory properties through the geometric property of these new SE manifolds.
There is an associated three dimensional log-terminal singularity $X$ for each SE manifold $L_X$, and for quasi-regular case, there is an associated two dimensional log Del Pezzo surface $(S,\Delta)$. The algebraic geometrical methods are quite useful in extracting interesting physical properties from singularity and log Del Pezzo surface.
The necessary and sufficient condition for the existence of SE metric on $L_X$ is related to K stability of $X$. Motivated by dual field theory, we propose a conjecture 
on how to reduce the check of K stability to possibly finite cases, which hopefully would give us a guideline to find a much larger space of SE metrics.  }

\begin{document} 
\maketitle
\flushbottom

\section{Introduction}
 AdS/CFT pairs can be  found using following method \cite{Maldacena:1997re,Klebanov:1998hh,Morrison:1998cs}: consider a string/M theory ${\cal T}$ on following background:
\begin{equation}
R^{1,d}\times X,
\end{equation}
here $X$ is  a local singularity which appears in the degeneration limit of a compact manifold with special holonomy $X^{'}$. The metric 
structure on $X^{'}$ would typically constrains the type of singularity $X$.  If we 
put N branes at the tip of $X$,  one expect to get a supersymmetric local field theory on the world volume of branes (the number of supersymmetries are 
determined by special holonomy type of $X^{'}$). 
If $X$ admits a conical metric of special holonomy:
(of the same type as $X^{'}$): 
\begin{equation}
g_X=dr^2+r^2 dg_{L_X};
\end{equation}
Here $L_X$ is a one dimensional lower manifold which is called the link of $X$; then we have a superconformal field theory (SCFT) on the brane which in the large N limit is dual to string/M theory ${\cal T}$ on following 
background 
\begin{equation}
AdS_{d+2}\times L_X. 
\end{equation}
There is no brane in this background, but one need to turn on corresponding flux of the branes \cite{Aharony:1999ti}. 
In this framework, the task of finding a AdS/CFT pair is reduced to find a singularity $X$ with conic metric of special holonomy, which can be 
thought of as the definition of SCFT. Many interesting properties of SCFTs can be learned from the geometric properties of $X$ and $L_X$.

In this paper, we take ${\cal T}$ to be type IIB string theory and $X$ to be a singularity of degeneration limit of a compact Calabi-Yau three manifold. 
$X$ is called three dimensional log terminal singularity (If the Calabi-Yau manifold is simply connected, one actually get a 3d canonical singularity.).  We put D3 branes on the tip of $X$ to get a four dimensional $\mathcal{N}=1$ supersymmetric field theory.
One crucial fact is that not every 3d log terminal singularity admits a Ricci-flat conical metric 
\footnote{They can admit other type of Calabi-Yau metric though.}. There are sufficient and necessary conditions about the existence of Ricci-flat conic metric on $X$ \cite{collins2015sasaki}:
\begin{itemize}
\item $X$ is an affine variety and admits an effective $C^*$ action such that the canonical top form $\Omega$ has charge two. 
\item $X$ is K stable with respect to above $C^*$ action. 
\end{itemize}   
The physical interpretation of K stability is given in \cite{collins2016k}. The link $L_X$ admits a positive Sasakian structure if $X$ is log terminal and has an effective $C^*$ action, and it admits a Sasaki-Einstein (SE) metric 
if $X$ is K stable. Four dimensional $\mathcal{N}=1$ SCFT has an important $U(1)_R$ symmetry which is identified with the above $C^*$ action of K stable singularity $X$. 
Therefore the classification of SCFT in this framework is reduced to find K stable log terminal singularity with a preferred $C^*$ action. 

In practice, it is not easy to check $K$ stability explicitly; but there are also sufficient theorems on the existence of Ricci-flat conic metric.  Examples 
include the toric SE manifold \cite{futaki2009transverse}, and some hypersurface examples \cite{boyer2005einstein}. With the K stability tools and the existence theorems, we 
do have a huge class of new SE manifolds which have not been studied physically before. 
Previously the physical application is mainly focusing on toric singularity \cite{Franco:2005sm}, and the main purpose of this paper is to initiate  a 
study of these new SCFTs defined by these general SE manifolds. These large class of examples provide a very interesting space of theories which
one can use to study  four dimensional $\mathcal{N}=1$ SCFT and AdS/CFT correspondence.  

It is in general difficult to write down explicit SE metrics on $L_X$ (see however \cite{gauntlett2004sasaki}  for some explicit metrics.), however what makes it possible to learn lots of interesting field theory 
properties is the connection of SE manifold to algebraic geometry. First, one have an associated algebraic object $X$ which is an affine singularity with a $C^*$ action, and some of nontrivial field theory properties 
are easily found from $X$:
 \begin{itemize}
 \item The affine ring of $X$ determines the chiral ring of the field theory in the large N limit. One can extract central charge from the Hilbert series of $X$ with respect to $C^*$ action.
 \item  The $C^*$ action is related to the $U(1)_R$ symmetry and the normalization is fixed by requiring canonical volume form $\Omega$ to have charge two.
 \item Some mesonic global symmetries can be read from automorphism group of the singularity. The homology group of the link $L_X$ can often be computed using the data on $X$ too, from which one can get more information of  field theory, such as the number of baryonic global symmetries. 
 \end{itemize}
Moreover, if the Sasakian structure on $L_X$ is quasi-regular, one have a $C^*$ fibration over a base surface which is a log Del Pezzo surface  $(S,\Delta)$ \cite{kollar2005einstein}. Such 
surfaces are studied thoroughly in mathematics literature, and algebraic geometrical methods are also quite useful in studying the field theory properties
such as the classification.

The connection between the field theory and singularity $X$ provides a physical interpretation of K stability \cite{collins2016k}, and physical arguments seem to suggest that
one only need to check finite number of specific test configurations. We propose an explicit conjecture of such reduction for hypersurface singularity, and 
if true it would give us a much larger space of SE manifolds.   

This paper is organized as follows: Section 2 reviews some basic facts about four dimensional $\mathcal{N}=1$ SCFT; Section 3
review basic properties of Sasaki-Einstein manifold; Section 4 describes several class of known five dimensional Sasaki-Einstein manifolds; Section 5 describes the 
map between the physical properties of field theory and the geometric aspect of $X$ and $L_X$; Section 6 gives a conjecture about reducing the check of K stability to finite cases;  
Finally a conclusion is given in section 7.

\section{Generalities of four dimensional $\mathcal{N}=1$ SCFT}

\textbf{Representation of  $\mathcal{N}=1$ superconformal algebra}: Four dimensional $\mathcal{N}=1$ superconformal algebra
has a bosonic symmetric group $SO(2,4)\times U(1)_R \times G$: here $SO(2,4)$ is four dimensional conformal group,  $U(1)_R$ is a 
$R$ symmetry group which acts non-trivially on supercharges, and $G$ is other global symmetry group. A highest weight representation of 
$\mathcal{N}=1$ superconformal algebra is labeled as $|\Delta, r, j, \bar{j}\rangle$, where $\Delta$ is the scaling dimension, $r$ 
is $U(1)_R$ charge and $j, \bar{j}$ are left and right spins.
If these quantum numbers satisfy certain conditions, the corresponding supermultiplet becomes short. 
The shortening conditions are studied in \cite{Osborn:1998qu} and is  summarized in table \ref{table:short}.
\begin{table}[!htb]
\begin{center}
  \begin{tabular}{ |c|c|c|c|}
    \hline
       ${\cal B}$ &$\bar{Q}_{\dot{\alpha}} | \rangle=0,,~~\bar{j}=0$&$\Delta={3\over 2}r$& ${\cal B}_{r,(j,0)}$  \\ \hline
              $\bar{{\cal B}}$ &$Q_{\alpha} | \rangle=0,,~~j=0$&$\Delta=-{3\over 2}r$& $\bar{{\cal B}}_{r,(0,\bar{j})}$  \\ \hline
              ${\cal C}$ & $(\bar{Q}_1+{1\over 2 \bar{j}}\bar{Q}_2\bar{J}_{-})| \rangle=0,~\bar{j}>0,~~\bar{Q}^2|\rangle=0,\bar{j}=0$ &$\Delta=2+2\bar{j}+{3\over 2}r$& ${\cal C}_{r, (j, \bar{j})}$ \\ \hline
                           $\bar{{\cal C}}$ & $ (Q_2-{1\over 2j}Q_1J_{-})| \rangle =0,~j>0,~~Q^2|\rangle=0,~j=0$ &$\Delta=2+2j-{3\over 2}r$& $\bar{{\cal C}}_{r, (j, \bar{j})}$ \\ \hline
     $\hat{{\cal C}}$ & ${\cal C}\cap {\cal \bar{C}}$ & $\Delta=2+j+\bar{j},~{3\over 2} r=j-\bar{j}$ & $\hat{{\cal C}}_{(j, \bar{j})}$ \\ \hline 
     ${\cal D}$&${\cal B} \cap \bar{{\cal C}}$ & $\bar{j}=0, {3\over 2}r=j+1, \Delta=1+j$ & ${\cal D}_{(j, 0)}$ \\ \hline
          $\bar{{\cal D}}$&$\bar{{\cal B}} \cap {\cal C}$ & $\bar{j}=0, -{3\over 2}r=\bar{j}+1, \Delta=1+\bar{j}$ & $\bar{{\cal D}}_{(0, \bar{j})}$ \\ \hline
  \end{tabular}
  \end{center}
  \caption{Various shorting condition for a highest weight representation of $\mathcal{N}=1$ superconformal algebra. }
  \label{table:short}
\end{table}
${\cal B}$ type operator is called chiral operator, and the unitarity bound implies the constraint on $r$ charge: $r\geq {2\over3}$; and $\hat{{\cal C}}$ multiplet contains conserved current, i.e. $\hat{{\cal C}}_{{1\over 2},{1\over 2}}$ contains the energy-momentum tensor 
and conserved $U(1)_R$ current, and $\hat{{\cal C}}_{0,{1\over 2}}$ contains  conserved current for other supersymmetries, and $\hat{{\cal C}}_{0,0}$ includes the current for other global symmetries.

\textbf{Superconformal index}:
Choose a supercharge $L=\bar{Q}_{\dot{1}}$, which satisfy the commutation relation
\begin{equation}
\{L,L^{+}\}=2{\cal H},~L^2=0.
\end{equation}
here ${\cal H}=H-2\bar{j}-{3\over 2}r$. Using the standard argument leading to Witten index \cite{witten1982constraints}, we can 
define superconformal index as the following trace \cite{Kinney:2005ej}:
\begin{equation}
I(x,t)=\text{Tr}(-1)^{F}t^{({2\over 3}\Delta+{4\over3}\bar{j})}x^{2J_3}.
\end{equation}
The trace only receive contribution  from the states with ${\cal H}=0$, so only states with $\Delta=2\bar{j}+{3\over2}r$ will contribute 
to this index. The multiplet that contributes to the index are ${\cal B}, {\cal C}, \hat{{\cal C}}, {\cal D},\bar{{\cal D}}$:
\begin{align}
&{\cal B}_{r,(j,0)}:~~I(x,t)=(-1)^{2j}t^r{\chi_{2j+1}(x)\over (1-tx)(1-tx^{-1})} \nonumber\\
&{\cal C}_{r,(j,\bar{j})}:~~I(x,t)=-(-1)^{2j+2\bar{j}}t^{2\bar{j}+2+r}{\chi_{2j+1}(x)\over (1-tx)(1-tx^{-1})} \nonumber\\
&\hat{{\cal C}}_{(j,\bar{j})}:~~~I(x,t)=-(-1)^{2j+2\bar{j}}t^{{2\over3}(j+2\bar{j}+3)}{\chi_{2j+1}(x)\over (1-tx)(1-tx^{-1})} \nonumber\\
&{\cal D}_{(j,0)}:~~I(x,t)=(-1)^{2j}t^{{2\over3}(j+1)}{\chi_{2j+1}(x)-t\chi_{2j}(x)\over(1-tx)(1-tx^{-1})} \nonumber\\
&\bar{{\cal D}}_{(0,\bar{j})}:~~I(x,t)=-(-1)^{2\bar{j}}{t^{{4\over3}(\bar{j}+1)}\over (1-tx)(1-tx^{-1})} 
\end{align}
Here $\chi_{2j+1}(x)=x^j+\ldots+x^{-j}$.  

\textbf{Chiral ring}:
Of particular interest are the chiral operators ${\cal B}$. They are defined as the operators such that $\{ \bar{Q}_{\dot{\alpha}}, {\cal O}\}=0$. The correlation function of these operators
\begin{equation}
\langle {\cal O}_1(x_1)\ldots {\cal O}_n(x_n)\rangle,
\end{equation}
is independent of positions \cite{cachazo2003chiral}. For a chiral operator, the addition of a $\bar{Q}_{\dot{\alpha}}$ commutator does not change the 
chiral correlation function, so by chiral operator we really mean the cohomology class of the operators 
\begin{equation}
{\cal O}\sim {\cal O}+\{\bar{Q}_{\dot{\alpha}},\chi\}.
\end{equation}
The OPE of chiral operators takes the following structure 
\begin{equation}
O_iO_j=\sum_k C_{ij}^k O_k.
\end{equation}
Here $C_{ij}^k$ does not depend on the position of insertion. 
So the chiral operators form a commutative ring with a unit, and is called chiral ring. The determination of chiral ring is quite useful in understanding 
the property of a SCFT, i.e. the moduli space of vacua.

One can also study the counting of chiral operators. We can define following generating function:
\begin{equation}
H(t)=\sum_\alpha t^{\alpha}\text{dim}H_\alpha;
\end{equation}
Here $H_\alpha$ is the space of chiral operator with $U(1)_R$ charge $\alpha$. The above generating function is not a protected quantity, but 
it could also contains important information of the field theory.

\textbf{Central charges}:
For a 4d CFT, one can define the central charges $a, c$ from the expectation value of the trace part of energy-momentum tensor:
\begin{equation}
\langle T^\mu_\mu \rangle_{g_{\mu \nu}}=-a {E\over 16 \pi^2}+c {W^2\over 16 \pi^2}.
\end{equation}
Here $E$ is Euler tensor and $W$ is the Weyl tensor of the background metric $g_{\mu\nu}$. It is often difficult to compute these central charges for 
a generic 4d CFT. However, for 4d $\mathcal{N}=1$ SCFT, the central charge is related to the anomaly of $U(1)_R$ symmetry as follows:
\begin{equation}
a={3\over 32}(3 tr R^3-tr R),~~c={1\over 32}(9 tr R^3-5 tr R),
\end{equation}
and the above formula  makes it possible to compute  central charges exactly by computing the anomaly of $U(1)_R$ symmetry.
In practice, we usually do not know the $U(1)_R$ symmetry and its anomaly of a 4d $\mathcal{N}=1$ SCFT. Intrilligator and Wecht proved that the correct $R$ symmetry maximize $a$ \cite{Intriligator:2003jj}. Namely, consider the linear combination of all anomaly free abelian symmetries of SCFT: 
\begin{equation}
R_s=\sum s_I F_I,
\end{equation}
then the correct $U(1)_R$ symmetry is determined by maximizing the  trial central charge $a(s_I)$.  So if we can determine all possible symmetries and its anomaly of the SCFT, one can also compute the central charges using $a$ maximization. In practice, a SCFT is often described as the IR fixed point of 
a UV theory with Lagrangian description, and one can do the a maximization using 
the symmetries and anomalies from UV Lagrangian description. However, one should always keep in mind that the above  method often \textbf{does not} include all possible symmetries of IR SCFT, since one could have accidental symmetries which is not visible from the UV Lagrangian.

\textbf{Deformations}:
Given a 4d $\mathcal{N}=1$ SCFT, it is interesting to study its supersymmetry preserving deformations. The relevant and marginal deformations were classified in \cite{Green:2010da}. 
A chiral operator ${\cal B}_{2,(0,0)}$ is a marginal operator since it has scaling dimension three. To determine whether it is exact marginal, one need to know its quantum property under other global symmetries.
It is proven in \cite{Green:2010da} that such an operator is exact marginal if it is a singlet under the global symmetry. 
The relevant deformations are associated with the operators ${\cal B}_{r,(0,0)}$ with ${2\over 3}\leq r<2$.

\textbf{Parameter $N$}: 
If is often possible to define a family of 4d $\mathcal{N}=1$ SCFTs parameterized by an integer $N$. In the large N limit, various properties of the SCFT can be simplified, i.e. the chiral ring might have a simpler description. Various operators can be separated into two kinds depending its behavior with respect to parameter $N$: they are called mesonic operator if its scaling dimension do not depend on $N$, and baryonic operator if it depends on some power of $N$. If our SCFT has an exact marginal deformation with deformation parameter $\tau$, we might consider a 't Hooft limit 
\begin{equation}
N\rightarrow \infty,~~N\tau~\text{fixed},
\end{equation}
and properties of the theory might be further simplified in this limit.

\section{Sasakian-Einstein Manifold}

\subsection{Sasakian manifold}
\subsubsection{Definition}
Here we review the basic aspect of Sasakian manifold, for more details, see \cite{boyer2008sasakian,sparks2010sasaki}.
We define the Sasakian structure on a $(2n+1)$ dimensional manifold $M$ by starting with a contact form $\eta$, which is a one form such that $\eta\wedge (d\eta)^n \neq 0$ everywhere on $M$.
There is a unique \textbf{Reeb} vector field $\zeta$ associated with it, and 
satisfying following important conditions:
\begin{equation}
i_\zeta(\eta)=1,~~i_\zeta(d\eta)=0.
\end{equation}
This Reeb vector field $\zeta$ plays a distinguished role in the study of Sasakian manifold. 
$\eta$ itself defines a $2n$ dimensional contact bundle $D$ whose fibre is defined as the space of vector fields $V$ satisfying $\eta(V)=0$. The tangent bundle then has a factorization $T_M=L_\zeta\oplus D$, where $L_\zeta$
is generated by vector field $\zeta$. We then define a complex structure on $D$, and extend it to a $(1,1)$ tensor $\Phi$ on the tangent bundle of $M$ such that 
$\Phi \zeta=0$. Once $\Phi$ is given, we get a metric $g=d\eta ( \Phi)+\eta\bigotimes \eta$. So a Sasakian manifold might be defined by following data:
\begin{equation}
\boxed{M(\zeta,\eta,\Phi,g)}
\end{equation}
with $g$ defined using $\Phi$ and $\eta$. Let's list some of the metric properties: 
\begin{align}
& g(\Phi X, \Phi Y)=g(X,Y)-\eta (X) \eta (Y),~~L_\zeta g=0,~L_\zeta \Phi=0, \nonumber\\
&R(X,Y)\zeta=\eta(Y)X-\eta(X)Y, \nonumber\\
&R(X,\zeta)Y=\eta(Y)X-g(X,Y)\zeta.
\end{align}
Here $L_\zeta$ is the Lie derivative with respect to $\zeta$.  
The first equation implies that $\zeta$ has unit length, and $\zeta$ is the Killing vector field. 
Alternatively, we can define Sasakian structure using the following equivalent characterizations:
\begin{enumerate}
\item There exists a Killing vector field $\zeta$ of unit length on $M$ so that the tensor field defined by $\Phi(X)=\Delta_X\zeta$ satisfies the following condition 
\begin{equation}
(\Delta_X\Phi)(Y)=g(\zeta, Y)X-g(X,Y)\zeta.
\end{equation}
\item There exists a Killing vector field $\zeta$ of unit length so that the Riemann curvature satisfies the condition
\begin{equation}
R(X,\zeta)Y=\eta(Y)X-g(X,Y)\zeta. 
\end{equation}
\item There exists a Killing vector field $\zeta$ of unit length on $M$ so that the sectional curvature of every section containing $\zeta$ equals one.
\item The metric cone over $M$ is Kahler. Here the metric cone of $M$  is $C(M) =R^{+} \times M$, and the conic metric is 
\begin{equation}
g=dr^2+r^2 dg_M^2.
\end{equation}
\end{enumerate}
The fourth definition is the most used one in the literature. Let's describe more details here. Recall that a Sasakian 
manifold is the data $M(\zeta, \eta, \Phi, g)$, and the tangent bundle of $C(M)$ is generated by the vectors $Y$ defined on $M$, 
and vector $\Psi=r{\partial \over \partial r}$. We define a complex structure $I$ on $C(M)$ as follows:
\begin{equation}
I(Y)=\Phi Y+\eta(Y) \Psi,~~I(\Psi)=-\zeta. 
\end{equation}
Using the first equation, we have $I(\zeta)=\Psi$.
Let's choose holomorphic coordinates $z_i$ on contact bundle $D$ (recall that the D is the sub-bundle of tangent bundle $TM$ whose 
fibre are vector fields satisfying $\eta(X)=0$.), so the holomorphic tangent bundle of $C(M)$ is generated by $({\partial \over \partial z_1},\ldots,{\partial \over \partial z_n},\Psi-i\zeta)$. 
$\Psi-i\zeta$ is a holomorphic vector field and $L_{\Psi+i\zeta} g= 2g$. Moreover, let's choose holomorphic coordinates $(z_1,\ldots, z_n, z)$ and 
consider the generator of canonical bundle: $\Omega=dz_1\wedge \ldots \wedge dz_n \wedge dz$, here $z$ is the dual holomorphic coordinate for the vector 
field $\Psi-i\zeta$. We then have {$L_{\Psi+i\zeta} \Omega= (n+1)\Omega$} \footnote{We use the fact that $\omega^{n+1}=\Omega\wedge \bar{\Omega}$, and $\omega$ is Kahler form which has
charge two under anti-holomorphic vector field $\Psi+i\zeta$.}.

Sasakian manifold can be further classified using the property of Reeb vector field $\zeta$.
Given a no-where vanishing vector field $V$ on a manifold $M$, one can define a one dimensional foliation, i.e. the leaves are 
one dimensional sub-manifold defined as the integral curve of $V$. For a Sasakian manifold, 
one can use Reeb vector field $\zeta$ to define a foliation. The leafs are the integral curve of  $\zeta$:
\begin{equation}
{d\vec{x}\over dt}=\zeta(\vec{x}),
\end{equation}
here $\vec{x}$ is the local coordinates on $M$. This 
foliation is called characteristic foliation and we denoted it as ${\cal F}_\zeta$, which  can be further classified as follows:
 $F_\zeta$ is called 
\textbf{quasi-regular} if there is a positive integer k such that each point has a foliated chart $U$ such that each leaf of $F_\zeta$ 
passes through $U$ at most $k$ times, otherwise we call it \textbf{irregular}. If $k=1$, $F_\zeta$ is further called \textbf{regular}.  We call the 
corresponding foliation by the property of $F_\zeta$. There are a couple of useful facts:
\begin{enumerate}
\item  There are at least \textbf{two} dimensional space of Killing vector fields for irregular Sasakian manifold \cite{boyer2008sasakian}.
\item  Every Sasakian manifold $M$ admits quasi-regular Sasakian structure \cite{boyer2008sasakian}. The basic
reasoning is follows: if we start with a irregular Sasakian structure, we can deform it to a quasi-regular Sasakian structure. 
\end{enumerate}

Given a foliation $F_\zeta$ on Sasakian manifold $M$, one can define a transverse geometry in the following way.  
Let $\{U_\alpha\}$ be an open covering of $M$ and $\pi^\alpha: U_\alpha \rightarrow V_\alpha \in C^n$ submersions such that when
$U_\alpha\cap \U_\beta \neq \emptyset$, 
\begin{equation}
\pi_\alpha \pi^{-1}_\beta : \pi_\beta(U_\alpha\cap \U_\beta) \rightarrow \pi_\alpha(U_\alpha\cap \U_\beta)
\end{equation}
is bi-holomorphic. On each $V_\alpha$, we can give a Kahler structure as follows. Let $D=\text{Ker}~\eta \in TM$, there is a canonical isomorphism 
\begin{equation}
d\pi_\alpha: D_p\rightarrow T_{\pi_\alpha(p)}V_\alpha
\end{equation} 
Since $\zeta$ generates the isometry, the restriction of Sasakian metric $g$ to $D$ gives a well defined Hermitian metric $g_{\alpha}^T$ on $V_\alpha$.
This metric is in fact Kahler and the proof is as follows. Let $z^1,\ldots, z^n$ to be the holomorphic coordinates on $V_\alpha$, and use the same 
letter as the pull-back coordinates on $U_\alpha$. Let $x$ be the coordinate along the leaves with $\zeta = {\partial \over \partial x}$, then $x,z^1,\ldots, z^n$ 
forms local coordinates on $U_\alpha$. $(D\oplus {\cal C})^{1,0}$ is generated by the vectors of the following form
\begin{equation}
{\partial \over \partial z^i}+a_i \zeta
\end{equation}
and $a_i=-\eta({\partial \over \partial z_i})$ so that $\eta({\partial \over \partial z^i}+a_i \zeta)=0$. Since $i_\zeta (d\eta)=0$, 
\begin{equation}
d\eta({\partial \over \partial z^i}+a_i \zeta,\overline{{\partial \over \partial z^j}+a_j \zeta})=d\eta({\partial \over \partial z^i},{\partial \over \partial z^j})
\end{equation}
Therefore the fundamental two form $\omega_\alpha$ of the Hermitian metric $g_\alpha^T$ on $V_\alpha$ is the same as the restriction 
of ${1\over 2} d\eta$ on slice $x=constant$ in $U_\alpha$.  Since the restriction of a closed two form to a sub-manifold is  closed in general, then $\omega_\alpha$ is
closed. By this construction  $\pi_\alpha \pi^{-1}_\beta$ gives an isometry of Kahler manifold. Thus the foliation defined is a transverse Kahler foliation.

A differential form $\omega$ on M is said to be \textbf{basic} for the foliation  $F_\zeta$ if the following conditions hold:
\begin{equation}
i_\zeta\omega=0,~~L_\zeta(\omega)=0.
\end{equation}
We use $\Lambda^p$ to denote the sheaf of basic p forms, and $\Omega^p$ to denote the space of global section of $\Lambda^p$.
In a local foliated coordinate with $\phi=(x;z_1,\ldots, z_n)$, a basic $(p,q)$ form $\omega$ takes the form
\begin{equation}
\omega=\omega_{i_1,\ldots,i_r,\bar{j}_1,\ldots, \bar{j}_q} dz^{i_1}\wedge \ldots dz^{i_p}\wedge d\bar{z}^{j_1}\wedge \ldots d\bar{z}^{i_q}
\end{equation} 
Let $\Lambda^{p,q}_B$ be sheaf of  basic $(p,q)$ forms, we thus have the well defined operators
\begin{align}
& \partial_B:~\Lambda^{p,q}\rightarrow \Lambda^{p+1,q},~~~\bar{\partial}_B:\Lambda^{p,q}\rightarrow \Lambda^{p,q+1}.
\end{align}
If we set $d_B=d|\Omega^p_B$, we have $d_B=\partial_B+\bar{\partial}_B$, let $d_B^c={i\over 2}(\bar{\partial}_B-\partial_B)$, we have $d_Bd_B^c=i\partial_B\bar{\partial}_B$ and 
$d_B^2=(d_B^c)^2=0$. Let $d_B^*$ be adjoint of $d_B$, we thus have the basic Laplacian 
\begin{equation}
\Lambda_B=d_Bd_B^*+d_Bd_B^*
\end{equation}
We then have the basic de Rham complex $(\Omega_B^*,d_B)$ and the basic Dolbeault complex $(\Omega^{p,*},\bar{\partial}_B)$ whose cohomology groups are called 
the basic cohomology groups.

In our case, we have three important basic forms: the first one is $d\eta$ and it is obviously satisfying the condition of basic form and it is a $(1,1)$ form; The first Chern class $c_1(D)$ of 
contact bundle D; and finally
the transverse Ricci form $Ricci_T$ (Remember the transverse metric is defined as $g_T=d\eta ( \Phi)+\eta\bigotimes \eta$). The
cohomology class of $Ricci_T$ is denoted as  the basic Chern class $c_1(F_\zeta)$.  Using $c_1(F_\zeta)$, we have

\textbf{Definition}: A Sasakian structure $S$ is said to be of \textbf{positive} (negative) type if $c_1(F_\zeta)$ can be represented by a \textbf{positive} (negative) definite 
$(1,1)$ form. If either of these two conditions is satisfied, $S$ is said to be of definite type. $S$ is said to be null type if $c_1(F_\zeta)=0$. A Sasakian structure $S$ 
is called anti-canonical (canonical) if $c_1(F_\zeta)$ is a positive (negative) multiple of $[d\eta]_B$.

\subsubsection{Log terminal singularity with $C^*$ action}
As we reviewed earlier, one way to define Sasakian structure on a $2n+1$ dimensional manifold $M$ is using the metric cone $C(M)=R^{+}\times M$. We now 
include the origin to $C(M)$ and the corresponding variety is an affine variety so we can use algebraic geometry method to study its property. 
We are interested in positive Sasakian manifold, and from algebraic geometry point of view,  the corresponding cone has following important properties \cite{kollar2005einstein}:
\begin{enumerate}
\item The cone plus an extra point at origin is a $n+1$ dimensional affine variety which can be defined by the following ring $X$:
\begin{equation}
X=C[x_1,\ldots, x_i]/I
\end{equation}
Here $I$ is an ideal which is defined by some polynomials $I=\langle f_1, \ldots, f_r \rangle$. 
\item There is an effective $C^*$ action (all the coordinates have positive charge on $X$, and $X$ has a singularity at origin.
\item The singularity is a log terminal singularity, and we will give a definition below.   

A variety $X$ is said to have \textbf{canonical} singularity if its is normal \footnote{Normal means that the singularity is at most co-dimension two, i.e., for a three dimensional singularity, the singular locus is at most one dimensional.} and the following two conditions are satisfied \cite{reid1980canonical}:
\begin{itemize}
\item the Weil divisor $K_X$ \footnote{$K_X$ is the canonical divisor associated with X.} is Q-Cartier, i.e. $rK_X$ is a Cartier divisor \footnote{A Cartier divisor implies that it can be used to define a line bundle.}. Here $r$ is called index of the singularity.
\item for any resolution
of singularities $f: Y\rightarrow X$, with exceptional divisors $E_i\in Y$, the rational numbers $a_i$ satisfying 
\begin{equation}
K_Y=f^{*} K_X+\sum a_i E_i.
\end{equation}
are nonnegative.
\end{itemize}
It is called \textbf{log terminal} if $a_i>-1$. Index one canonical singularity is also called \textbf{rational Gorenstein singularity}.
\end{enumerate}
Log terminal singularity is always rational, and if it is Gorenstein (index is one), then it has to be canonical. Rational Gorenstein singularity is always canonical singularity.
Rational Gorenstein singularity plays a special role because any log terminal singularity is just a $Z_n$ quotient of a rational Gorenstein singularity. In general, however, not every abelian $Z_n$ quotient of rational Gorenstein singularity would give us a log terminal singularity, and it is also not known 
which abelian finite group of a rational Gorenstein singularity would give us a log terminal singularity.

Two dimensional log terminal singularity is classified by quotient singularity $C^2/G$ with $G$ a finite subgroup of $U(2)$; and 
if $G\subset SU(2)$, then it gives all the two dimensional rational Gorenstein singularity. 

There is no complete classification of three dimensional log terminal singularity, but we do know a large class of explicit examples of rational Gorenstein singularity. 
To a rational Gorenstein 3-fold singularity, one can attach a natural integer $k\geq 0$, such that:
\begin{itemize}
\item $k=0$, the singularity is a compound Du Val (cDV) point, i.e. the singularity can be written as follows 
\begin{equation}
f_{ADE}(x,y,z)+tg(x,y,z, t)=0;
\end{equation}
Here $f_{ADE}(x,y,z)$ denotes two dimensional Du Val singularity:
\begin{align}
& A_N:~~x^2+y^2+z^{N+1}=0,~~D_N:~~x^2+y^{N-1}+yz^{2}=0,~~E_6:~~x^2+y^3+z^{4}=0   \nonumber \\
& E_7:~~x^2+y^3+yz^{3}=0,~~E_8:~~x^2+y^3+z^{5}=0 \nonumber\\
\end{align}
If we further require that the singularity is \textbf{isolated} and has a $C^*$ action, the complete list of  such singularity has been found in \cite{wang2016classification}, see table. \ref{table:isolatedsingularitiesALEfib}.
\begin{table}[!htb]
\begin{center}
  \begin{tabular}{ |c|c| }
    \hline
       $J$& Singularity   \\ \hline
     $A_{N-1}$ &$x_1^2+x_2^2+x_3^N+z^k=0$ \\ \hline
 $~$& $x_1^2+x_2^2+x_3^N+x z^k=0$ \\ \hline
      
 $D_N$   & $x_1^2+x_2^{N-1}+x_2x_3^2+z^k=0$ \\     \hline
  $~$   &$x_1^2+x_2^{N-1}+x_2x_3^2+z^k x_3=0$ \\     \hline

  $E_6$  & $x_1^2+x_2^3+x_3^4+z^k=0$   \\     \hline
   $~$  & $x_1^2+x_2^3+x_3^4+z^k x_3=0$   \\     \hline
  $~$  & $x_1^2+x_2^3+x_3^4+z^k x_2=0$   \\     \hline

   $E_7$  & $x_1^2+x_2^3+x_2x_3^3+z^k=0$   \\     \hline
      $~$  & $x_1^2+x_2^3+x_2x_3^3+z^kx_3=0$     \\     \hline

    $E_8$   & $x_1^2+x_2^3+x_3^5+z^k=0$   \\     \hline
        $~$   & $x_1^2+x_2^3+x_3^5+z^k x_3=0$   \\     \hline
    $~$   & $x_1^2+x_2^3+x_3^5+z^k x_2=0$   \\     \hline

  \end{tabular}
  \end{center}
  \caption{3-fold isolated cDV singularities with $C^*$ action.  }
  \label{table:isolatedsingularitiesALEfib}
\end{table}

\end{itemize}

\begin{itemize}
\item If $k\geq 2$, then $k=mult_P X$. 
\item If $k\geq 3$, then $k+1$ is equal to the embedding dimension $=dim (m_p/m_p^2)$.
if $k=2$, then $P\in X$ is isomorphic to a hypersurface given by $x^2+f(y,z,t)$ with $f$ a sum of monomials with degree bigger than 4. If $k=1$, then 
then $P\in X$ is isomorphic to a hypersurface given by $x^2+y^3+f(y,z,t))=0$ with $f(y,z,t)=yf_1(z,t)+f_2(z,t)$ and $f_1$ (respectively $f_2$) is a sum of 
monomials $z^a t^b$ of degree $\geq 4$ (respectively $\geq 6$).
\item If $k=3$, the singularity $P$ is given by a hypersurface.
\item If $k= 4$, then $P$ is  equivalent to a complete intersection. 
\end{itemize} 
We then have following three class of log terminal examples:
\begin{enumerate}
\item Complete intersection singularity (They are Gorenstein). The rational condition implies that the maximal embedding dimension is $5$, so such singularities 
are defined by two polynomials. Hypersurface 
singularity and complete intersection rational singularity with a $C^*$ action has been classified in \cite{yau2005classification,Chen:2016bzh}, see
appendix. \ref{app: hyper} for the hypersurface example (we do not list the rational constraints on the parameters in the polynomial here, see \cite{yau2005classification}.).
\item The quotient singularity $C^3/G$ with $G$ a finite subgroup of $SU(3)$, and in fact these singularities are rational Gorenstein. 
\item Another class is Q-Gorenstein toric singularity, which is defined combinatorially by a convex polygon.  
\end{enumerate}

This cone perspective is quite useful as we can use many tools from algebraic geometry to study them. 
For example, the classification is reduced to the classification of log terminal singularity with a $C^*$ action. Many of these singularities have been studied in the context of four dimensional $\mathcal{N}=2$ SCFT in \cite{Xie:2015rpa,yau2005classification}.
The Sasakian manifold $M$ is defined as the link of $X$ \cite{looijenga1984isolated}.
If we require that $M$ is smooth, then $X$ is required to have 
\textbf{isolated} singularity.

In summary, one can get a Sasakian manifold $L_X$ from a $n+1$ dimensional log terminal singularity $X$ with a $C^*$ action, and $X$ is an affine variety:
\begin{equation}
\boxed{X=C[x_1,\ldots, x_i]/I}.
\end{equation} 
This ring is Q-Gorenstein and rational, and has canonical top form $\Omega$ which has definite weight under $C^*$ action: $L_\zeta \Omega=(n+1) \Omega$.

\subsubsection{Log Del Pezzo surface}\label{orbi}
There is another way of using algebraic geometry to study positive Sasakian manifold (We focus on Sasakian five manifold in this part).
Sasakian manifold has a Reeb vector field $\zeta$, and it always admit a quasi-regular Sasakian structure. 
This implies that the corresponding affine ring $X$ has 
a rational $C^*$ action. The above properties make it possible to associate 
a projective variety $S=X/ \{0\}/C^*$ with cyclic quotient singularity. There is also a natural $Q$ divisor $\Delta= \sum (1-{1\over m_i}) D_i$ on $S$, where $D_i$ is a divisor such that the stabilizer of $C^*$ action on $D_i$ has order $m_i$. The positivity of the Sasakian structure is equivalent to following condition on $S$:
\begin{equation}
-(K_S+\Delta)~~\text{is ample}.
\end{equation}
Here ample means that $-(K_S+\Delta)\cdot C>0$ for any irreducible curve $C$.
Given a pair $(S, \Delta)$, the Sasakian structures are specified by the following data:
\begin{enumerate}
\item Integers $0\leq b_i <m_i$ and $b_i$ is co-prime with $m_i$.
\item A class of Weyl divisor $B$.
\end{enumerate}
From these data on $S$, one can define a  ring and recover $X$:
\begin{align}
& R(B,\sum {b_i\over m_i})=\sum_j{\cal O}_X(jB+\sum [{jb_i\over m_i}]D_i), \\ \nonumber
& X= Spec_S R(B,\sum {b_i\over m_i}).
\end{align}
The map $X\rightarrow (S, \Delta)$ is called a Seifert structure \cite{kollar2005einstein}. 
The Chern class of the Seifert structure is defined as
\begin{equation}
CS(X/S)= B + \sum_i{b_i\over m_i}.
\end{equation}
Given a pair $(S,\Delta)$, we can define the following integer numbers:
\begin{align}
& m(x,\Delta)=\text{lcm}(m_i|x\in D_i), \\ \nonumber
& m(\Delta)=lcm(m_i).
\end{align}
These number plays a crucial role in determining the smooth Seifert structure: we have a \textbf{smooth} Seifert structure if $m(x,\Delta)*C(X/S)$ is a generator of local class group at $x$. 

So one can get a positive Sasakian manifold from following data:
\begin{equation}
\boxed{(S, \Delta=\sum_i (1-{1\over m_i})D_i),B+\sum_i{b_i\over m_i}D_i)}
\end{equation}
Here $B$ is a Weyl divisor, $0\leq b_i< m_i$ and is coprime with $m_i$, $-(K_S+\sum_i (1-{1\over m_i})D_i)$ is ample. Smooth of $M$
gives further constraint on $b_i$ and $B$.

\subsection{Sasaki-Einstein manifold}

We are now looking for a $(2n+1)$ dimensional Sasakian manifold which is also Einstein 
\begin{equation}
R_{ij}=\lambda g_{ij}.
\end{equation}
We are interested in the case where $\lambda$ is positive, and in this case, the constant is equal to $2n$. 
This is equivalent to the following condition
\begin{enumerate}
\item The conic metric $g=dr^2+r^2dg_M$ is Ricci flat.
\item In the quasi-regular case, it implies that the orbifold Kahler metric on the base is Kahler-Einstein with 
constant curvature $4n(n+1)$, and the Einstein constant is $2n+2$. 
\item The transverse Kahler metric $g_T$ is Einstein with Einstein constant $(2n+2)$.
\end{enumerate}
Let's fix the Reeb vector field $\zeta$, and the space of Sasaki-Einstein metric with fixed $\zeta$
might be called the moduli space of Sasaki-Einstein metric. Little is known about this space though. 

\subsubsection{Comments on spin structure}
Not all Sasakian manifold admits spin structure, and there is also Sasaki-Einstein manifold which does not have 
a spin structure. For a simply connected SE manifold, there is always a spin structure though.

One might wonder whether these non-spin SE manifold can play any role in string theory. In fact, they are also quite 
useful, and an important example is $RP^5$ whose associated cone is $C^3/Z_2$ with $Z_2$ action acts as reflection $(-1,-1,-1)$,
so $Z_2$ is not a finite subgroup of $SU(3)$. $RP^5$ does not admit spin structure, but it is quite useful to describe
the gravity dual of $SO(N)$ and $Sp(N)$ $\mathcal{N}=4$ SYM theory. 
Although 
not every Sasakian manifold admits spin structure, they all admit $Spin^c$ structure \cite{moroianu1997parallel}, so 
one could still work with spinors. Since in the $RP^5$ case, one need to study the orientifold of IIB string theory,
and also the discrete torsion of $B$ field is important. 
We also expect that one need to study orientifold and discrete torsion for these non-spin SE manifold. A useful observation is that non-simply connected SE manifold 
are derived from abelian quotient of 
rational Gorenstein singularity and therefore admit a finite abelian group action. It is interesting to further study IIB string construction 
on these non-spin SE manifolds.

A simply connected Sasaki-Einstein manifold admits at least two real Killing spinor, and the cone $X$ admits two parallel spinors.

\subsection{K stability and Sasaki-Einstein metric}
We focus on Sasakian five manifold in this section. Given a log terminal singularity $X$ with a $C^*$ action $\zeta$, 
the question is to determine whether the link $L_5$ has Sasaki-Einstein metric. This question is reduced to studying the K-stability of the ring $X$ \cite{collins2015sasaki}. 
In this subsection, we will discuss two crucial ingredients of K-stability;  test configuration and the Futaki invariant. The physical 
interpretation of K stability has been given in \cite{Collins:2016icw}.  K stability works for all log terminal singularity, not necessarily isolated singularity, in fact, in many cases it is crucial to 
consider test configuration involving non-isolated singularity. 
\subsubsection{New ring with extra $C^*$ action}
\begin{figure}[h]
\centering
  \includegraphics{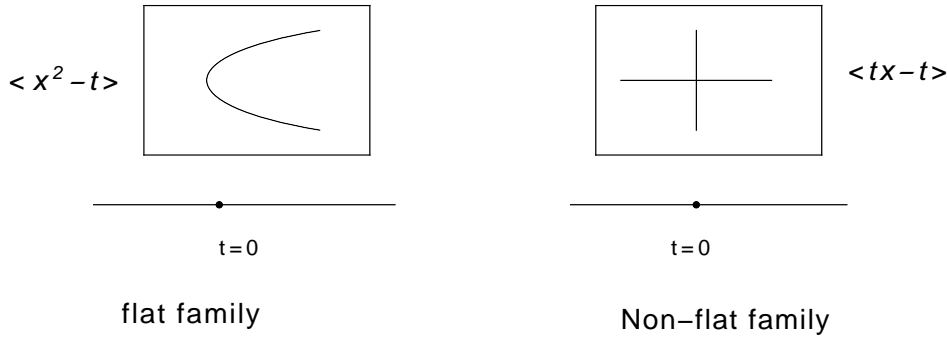}
  \caption{Left: A flat family of rings. At $t\neq 0$, there are two points and the configuration degenerates into one point at $t=0$, which is the central fibre of this flat limit. Right: A non-flat family of rings. At $t\neq 0$, the ring is zero dimensional, but at $t=0$, the ring is one dimensional. }
  \label{flat}
  \end{figure}

Let's first describe the definition of a test configuration arising in K stability. Let's start with a 3-fold log terminal singularity with a $C^*$ action. 
In the K-stability context, one constructs a test configuration by constructing 
a flat family $\pi: {\cal X} \rightarrow \mathbb{C}$ (for a simple illustration of flat and non-flat family, see figure. \ref{flat}.). This flat family is generated by a one dimensional symmetry generator $\eta$, and for $t\neq 0$, the ring  corresponding to the fiber $X_t = \pi^{-1}(t)$ is isomorphic to the original ring $X$. At $t=0$, the ring might degenerates into 
a different ring which we call ${X_0}$, and it is also called central fibre of a test configuration. 

The flat limit is a quite common concept in algebraic geometry, but its definition is quite involved and we do not want to give a detailed introduction here.  Here, we just want to  point out  several important features of the flat family constructed above.
\begin{itemize}
\item[(a)] The Hilbert series is not changed if we use the same symmetry generator for the new ring ${X_0}$.  In particular,  
$X_0$ has the same dimension as $X$.
\item[(b)] The maximal torus in the automorphism group of the central fibre $X_0$ has one more dimensional symmetry generated by $\eta$, unless $X_0 \cong X$. 
\end{itemize}
We require that the degeneration is normal (which implies that the codimension of the singular locus is at less than two). The new singularity $X_0$ in the non-trivial case, possesses an extra one-dimensional symmetry. 

\textbf{Example}: Consider the ring $X$ defined by the ideal $x^2+y^2+z^2+w^k=0$, and consider a $\mathbb{C}^*$ action $\eta$ which acts only on coordinate $w$ with the action $\eta(w)=t w$. We then get a family of 
rings parametrized by the coordinate $t$:
\begin{equation}
x^2+y^2+z^2+t^kw^k=0.
\end{equation} 
The flat limit of this family over $t=0$ is found (in this case) by keeping the terms with lowest order. The central fiber of this test configuration is then cut out by the equation
\begin{equation}
x^2+y^2+z^2=0.
\end{equation}
Notice that $l\eta$ with $l>0$ gives the same degeneration limit $X_0$. On the other hand $l\eta$ with $l<0$ gives a different degeneration limit-- we get the ring generated by the ideal $w^k=0$, which is not normal!

\subsubsection{Futaki invariant}
Now let's start with a ring $X$ with a Sasakian $C^*$ action $\zeta$, and we also choose the generators  $t_i, i=1,\ldots, n$ for the Lie algebra $\mathfrak{t}$ of the maximal torus in the automorphism group $G$ of X. We also have a canonical top form $\Omega$ which is the section of the sheaf associated with canonical divisor $K_X$, and 
this section is chosen to have definite weight under $C^*$ action. 
Let us write 
$\zeta=\sum_{i=1}^n \zeta_it_i$.
Consider a test configuration ${\cal X}$ generated by a symmetry generator $\eta$ and let $X_0$ denote the central fibre. We would like to determine whether or not $X_0$ destabilizes $X$. The 
crucial ingredient is Donaldson-Futaki invariant defined in \cite{donaldson2002scalar}. 

The ring $(X_0, \zeta, \eta)$ is still log terminal, and has a at least two dimensional symmetry group generated by $\zeta$ and $\eta$. We impose the following two conditions:
\begin{itemize}
\item[(a)] The charge on the coordinates $x_i$ is positive.
\item[(b)] The  $(3,0)$ form $\Omega$ has charge 2 \footnote{The holomorphic Reeb vector field has charge three on $\Omega$, but here we use a normalization so that the charge is the $U(1)_R$ charge. }.  
\end{itemize}
Hilbert series on ring $X_0$ with respect a $C^*$ action $x \zeta+ y\eta$ is defined as 
\begin{equation}
H(t)=\sum t^{\alpha(x \zeta+ y\eta)} dim H_{\alpha(x \zeta+ y\eta)}.
\end{equation}
here $H_\alpha$ is the subspace of ring $X$ with charge $\alpha$ under action $x\zeta+y\eta$. If we take $t=e^{-s}$, and we have the following expansion \cite{stanley1978hilbert}:
\begin{equation}
H(e^{-s})={a_0(x\zeta+y\eta)\over s^3}+{a_1(x\zeta+y\eta)\over s^2}+\ldots
\end{equation}
The second condition can be fixed by computing the Hilbert series of $X_0$ with respect to symmetry generator and imposing the condition $a_0=a_1$, so
we get a one dimensional space of Reeb vector field. $a_0$ is proportional to the volume of the link $L_X$. Let's try following parameterization of 
one dimensional Reeb vector field:
\begin{equation}
\zeta(\epsilon)=\zeta+\epsilon(\eta-a\zeta).
\end{equation}
Notice that we require $\epsilon>0$ so that the central fibre is the same as the original one if we use the symmetry $\epsilon(\eta-a\zeta)$ to generate the test configuration. 
Substitute the above parameterization into the equation $a_0=a_1$ (computed using new ring $X_0$) and expand it to first order in $\epsilon$,  we have
\begin{equation}
a_0(\zeta+\epsilon(\eta-a\zeta))=a_1(\zeta+\epsilon(\eta-a\zeta)) \rightarrow a_0(\zeta)+\epsilon (\eta-a \zeta)\cdot a_0^{'} = a_1(\zeta)+\epsilon (\eta-a \zeta)\cdot a_1^{'}.
\end{equation}
Here $a_0^{'}$ and $a_1^{'}$ are the vectors defined by the derivative ${da_i(\vec{x})\over d {\vec{x}}}|_{\vec{x}=\zeta}$, and $\vec{x}=\sum_{i=1}^ns_i t_i+b \eta$. Using the result $a_0(\zeta)=a_1(\zeta)$, we have 
\begin{equation}
a={\eta\cdot (a_0^{'}-a_1^{'}) \over \zeta \cdot (a_0^{'}-a_1^{'})}={\eta\cdot (a_1^{'}-a_0^{'}) \over a_0}={1\over a_0(\zeta)}({da_1(\zeta+\epsilon \eta) \over d \epsilon }-{d a_0(\zeta+\epsilon \eta) \over d\epsilon })|_{\epsilon=0}.
\label{constant}
\end{equation} 
We also use the fact $\zeta\cdot a_0^{'}=3 a_0(\zeta),~\zeta\cdot a_1^{'}=2a_1(\zeta)=2a_0(\zeta)$. 
Now the Futaki invariant is defined to be 
\begin{equation}
F(X,\zeta, \eta)=D_\epsilon a_0(\zeta(\epsilon))|_{\epsilon=0}.
\end{equation}
This definition is not of the form of the original Futaki invariant defined in \cite{donaldson2002scalar}, however, we will now show that our definition is equivalent to the original one (see also \cite{collins2015sasaki} for more discussion). We have
\begin{align}
& F(X,\zeta,\eta)=D_\epsilon a_0(\zeta+\epsilon(\eta-a\zeta))=(\eta-a \zeta) \cdot a_0^{'} \nonumber\\
&=D_\epsilon a_0(\zeta+\epsilon \eta)-{\eta\cdot (a_0^{'}-a_1^{'}) \over a_0}D_\epsilon a_0(\zeta+\epsilon \zeta) \nonumber\\
&=D_\epsilon a_0(\zeta+\epsilon \eta)+3a_0(\zeta){\eta\cdot (a_0^{'}-a_1^{'}) \over a_0} \nonumber \\
&= D_\epsilon a_0(\zeta+\epsilon \eta)+3a_0(\zeta)D_{\epsilon}{a_1(\zeta+\epsilon \eta)\over a_0(\zeta+\epsilon \eta)}. 
\label{futaki}
\end{align}
We use the definition from first line to second line, and from second line to third line we use the fact $D_\epsilon a_0(\zeta+\epsilon \zeta)=-3 a_0(\zeta)$ (which can be found using the definition of Hilbert series). 
The formula in the last line is precisely the Futaki invariant defined in \cite{collins2015sasaki}.  Having defined the Futaki invariant, we can now state the definition of K-stability.

\begin{theorem}
A polarized ring $(X,\zeta)$ is stable if for any non-trivial test configuration generated by the symmetry $\eta$, the Futaki invariant satisfies
\begin{equation}
F(X, \zeta, \eta)>0.
\end{equation}
And for the trivial test configuration, namely the central fibre $X_0$ is the same as $X$, the Futaki invariant satisfies
\begin{equation}
F(X, \zeta, \eta)\geq0.
\end{equation}
\end{theorem}
\begin{figure}[h]
\centering
  \includegraphics{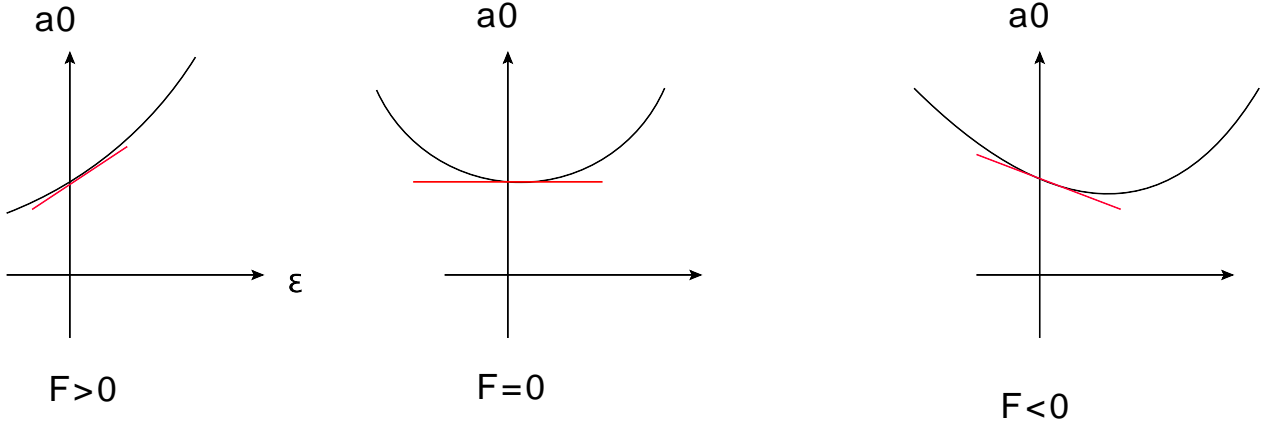}
  \caption{Three situations for Futaki invariant. $F>0$: $a_0(\epsilon)>a(0)$ for $\epsilon>0$; $F=0$: the minima of $a_0$ is achieved at $\epsilon=0$; $F<0$: the minima of $a_0$ is  achieved for $\epsilon>0$. 
   Notice that we only need to look at $a_0$ for $\epsilon>0$.}
  \label{futaki}
  \end{figure}
See figure. \ref{futaki} for the behavior of $a_0(\epsilon)$ with respect to $\epsilon$, and it is clear that if $F<0$, minimal of $a_0$ is achieved 
for $\epsilon>0$, which implies that the volume of $X_0$ is smaller than $X$. So K stability can be interpreted as the generalized volume minimization.

\textbf{Example}: Consider the ring $X$ which is generated by the ideal $x^2+y^2+z^2+w^k=0$, this ring has a symmetry $\zeta$ with charge $({2k\over k+2},{2k\over k+2},{2k\over k+2},{4\over k+2})$ on coordinates $(x,y,z,w)$. This symmetry is chosen such that the 
$(3,0)$ form $\Omega={dx \wedge dy \wedge dz \wedge dw\over df}$ has charge two. The Hilbert series for $\zeta$ is 
\begin{equation}
Hilb(X,\zeta, t)=\frac{1-t^{\frac{4 k}{k+2}}}{\left(1-t^{\frac{4}{k+2}}\right) \left(1-t^{\frac{2 k}{k+2}}\right)^3}.
\end{equation}
Expand around $t=1$, we find $a_0(\zeta)=a_1(\zeta)={(2+k)^3\over 8k^2}$. 
Now consider the test configuration generated by the symmetry $\eta$ with charges $(0,0,0,1)$.  In this case, the central fibre $X_0$ is generated by the
ideal $x^2+y^2+z^2=0$. Using formula \eqref{constant}, 
the one parameter possible $U(1)_R$ symmetry is 
\begin{equation}
\zeta(\epsilon)=\zeta+\epsilon(\eta-{1\over 2}\zeta).
\end{equation}
The Hilbert series with respect to above symmetry is 
\begin{equation}
Hilb(X_0,\zeta(\epsilon),t)={1-t^{(1-{\epsilon\over 2}){4k\over k+2}}\over (1-t^{(1-{\epsilon\over 2}){2k\over k+2}})^3(1-t^{(1-{\epsilon\over 2}){4\over k+2}+\epsilon)})}.
\end{equation}
Substituting $t=exp(-s)$ and expand the Hilbert series around $s=0$, we get 
\begin{equation}
a_0(\zeta(\epsilon))=a_1(\zeta(\epsilon))=\frac{2 (k+2)^3}{(\epsilon-2)^2 k^2 (\epsilon k+4)}.
\end{equation}
The Futaki invariant is computed as 
\begin{equation}
F=D_\epsilon a_0(\zeta(\epsilon))|_{\epsilon=0}=\frac{(4-k) (k+2)^3}{32 k^2}.
\end{equation}
So $F\leq 0$ for $k\geq 4$.  Since $X_0$ is clearly not isomorphic to $X$, we conclude that $X_0$ destabilizes $X$ for $k\geq 4$.  

\subsubsection{Some discussions}
Checking K-stability involves two steps.  First, finding a test configuration and then computing the Futaki invariant.  While the computation of Futaki invariant is straightforward, the set of possible test configurations is in principle infinite.  Thus, in order to check K-stability one needs to reducing the 
sets of possible test configurations. There are several simplifications we can make:
\begin{itemize}
\item The first simplification has already been used, namely we require that the central fibre to be normal. 
\item Assume that the symmetry group of the ring $X$ is $G$, then one only need to consider the flat families generated by a symmetry which commutes with $G$ \cite{datar2015k, collins2015sasaki}. This fact is quite useful 
for singularities with many symmetries. In particular, if the variety has three dimensional symmetries (or in other words, $X$ is toric), then there are no non-trivial test configurations, and hence checking stability reduces to volume minimization (or $a$-maximization). 
\end{itemize}

\section{Examples}
In this section, we are going to study several class of isolated three dimensional rational Gorenstein singularity with a $C^*$ action. The link of such 
singularity will carry a positive Sasakian structure. We then study the constraints such that there is a SE metric. To begin with, we will first review the topological 
constraint  which obstructs the existence of positive Sasakian structure on simply connected five manifold. We then study the link of  toric singularity, 
complete intersection singularity, quotient singularity and finally study five dimensional SE manifold from two dimensional log del Pezzo surface point of view.

\subsection{Simply connected Sasakian five manifold}
In general, it is not possible to classify five manifold under diffeomorphism. However, 
simply connected five manifold has been given a  classification by Smale and Barden \cite{boyer2008sasakian}.  They are specified by torsion subgroup $H_2(M_5,Z)$ and an invariant $i(M)$, 
see table. \ref{table:simply} for the basic building block. $i(M)=0$ implies that the manifold is a spin manifold.
The class $B$ of simply connected, closed, smooth, 5-manifolds is classifiable under diffeomorphism. Furthermore, any such $M$ is diffeomorphic to one 
of the spaces:
\begin{equation}
M_{j;k_1,\ldots, k_s}=X_j\# M_{k_1} \# \ldots \# M_{k_s},
\end{equation}
where $-1\leq j \leq \infty$,  $1<k_1$ and $k_i$ divides $k_{i+1}$ or $k_{i+1}=\infty$.  Here

\begin{table}[!htb]
\begin{center}
  \begin{tabular}{ |c|c|c|}
    \hline
       $M$ & $H_2(M,Z)$ & $i(M)$ \\ \hline
       $M_{-1}=SU(3)/SO(3)$&$Z_2$&1 \\ \hline
       $M_1=X_0=S^5$& $0$ & $0$ \\ \hline
       $X_j,0<j<\infty$ & $Z_{2j}\oplus Z_{2j}$ & j \\ \hline
       $X_\infty$ & Z & $\infty$ \\ \hline
       $M_k,1<k<\infty$ & $Z_k\oplus Z_k$ & 0 \\ \hline
       $M_\infty =S^2\times S^3$ & $Z$ & 0 
        \\ \hline
  \end{tabular}
  \end{center}
  \caption{ Basic simply connected five manifolds. }
  \label{table:simply}
\end{table}

The torsion subgroup of these simply connected manifolds which admit positive Sasakian structure is constrained. 
This result is proven by Kollar \cite{kollar2005einstein} using minimal model program. Recall that each quasi-regular positive Sasakian five manifold 
gives a two dimensional log del Pezzo surface $(S, \sum (1-{1\over m_i})D_i)$, here $S$ has cyclic quotient singularity. 
Such S is rational and one can use minimal model program to constrain the topology of $D_i$. The minimal model program goes as follows:
we first start with a log Del Pezzo surface $S$ and resolve the singularity, we then do the minimal model program and get a 
minimal rational surface which are $P^2, P^1\times P^1$ and Hirzbruch surfaces $F_n$. After a case by case analysis, 
it is proven in \cite{kollar2005einstein} that:
\begin{enumerate}
\item There is only one $D_i$ with $g(D_i)\geq 1$. Assume $D_0$ has genus bigger than 0 and $(1-{1\over m_0})\geq {1\over 2}$, then we have 
\begin{align}
& g(D_0)=1~if~(1-{1\over m_0}) \geq {5\over 6}, \nonumber \\
&g(D_0)=2~if~(1-{1\over m_0}) \geq {3\over 4}, \nonumber \\
& g(D_0)=4~if~(1-{1\over m_0}) \geq {2\over 3}.
\end{align}
\item One can compute the homology and cohomology group of the five dimensional Sasakian manifold $M$ from the data on the base $(S,\Delta)$. Let $f:M^5\rightarrow (S, \Delta=\sum_i(1-{1\over m_i})D_i)$ to be a smooth Seifert bundle over a projective surface with cyclic quotient singularity. Set $s=b_2(S)$ and assume that $H_1^{orb}(S,Z)=0$ \footnote{The pair $(X,\Delta)$ can be 
interpreted as an orbifold.
For an orbifold $(X,\Delta)$, the orbifold fundamental group $\pi_1^{orb}$   is the fundamental group of $X / (SingX \cup Supp \Delta)$
modulo the relations: if $\gamma$ is any small loop around $D_i$ at a smooth point then
 $\gamma^{ m_i} =1$. The abelianization of $\pi_1^{orb}$ is denoted by $H_1^{orb}(X,\Delta)$, called the abelian
orbifold fundamental group.} \footnote{Let S be a normal, projective surface with rational singularities. Then $H_1^{orb}(X,\Delta)=0$ if and only if 
\begin{itemize} 
\item $H_1(S^0,Z)=0$, here $S^0$ is the smooth part of $S$.
\item The map $H_2(S,Z)\rightarrow \sum_j Z/m_j$ given by $L\rightarrow (L\cdot D_j)$ mod $m_j$ is surjective.
\end{itemize}}. Then the integral cohomology groups $H^i(M^5,Z)$ are
 \begin{center}
  \begin{tabular}{ |c|c|c|c|c|c|c|}
    \hline
    i & 0 & 1&2 &3 &4 &5  \\ \hline
    $H^i(M,Z)$&Z&0&$Z^{s-1}\oplus Z_d$&$Z^{s-1}\oplus \sum_i{(Z_{m_i})}^{2g(D_i)}$ &$Z_d$ &Z \\ \hline
  \end{tabular}
\end{center}
here $d$ is the largest natural number such that $m(\Delta)c_1(Y/S) \in Cl(S)$ is divisible by $d$ (see the definitions in section \ref{orbi}), and $g(D_i)$ is the genus of divisor $D_i$.  One can find the homology group by looking at the dual of above cohomology groups. Because of the constraint presented in 1, the torsion subgroup $H_2(M,Z)$ is one of the following:
\begin{equation}
(Z_m)^2,~~(Z_5)^4,~~(Z_4)^4,~~(Z_3)^4,~(Z_3)^6,~(Z_3)^8,~(Z_2)^{2n}.
\end{equation}
\end{enumerate}
The torsion subgroup of non-simply connected Sasakian five manifold is further studied in \cite{kollar2009positive}.

\subsection{Toric singularity}
Let's start with a tree dimensional standard lattice $N$, and its dual lattice $M=Hom(N,Z)$. A rational convex polyhedral
 cone $\sigma$ is defined by a set of lattice vectors $\{ v_1,\ldots, v_n\}$ as follows:
\begin{equation}
\sigma=\{  r_1 v_1+r_2 v_2+\ldots+ r_n v_n | r_i\geq 0\}
\end{equation}
Its dual cone $\sigma^{\vee}$ is defined by the set in $M_R$ such that 
\begin{equation}
\sigma^{\vee}=\{ \langle w|v\rangle\geq 0|w\in M_R, v\in \sigma \}. 
\end{equation}
here we use the standard pairing between two lattices $N$ and $M$. A ray generator $v_\rho$ of $\sigma$ is a lattice vector,  such that 
it is not a multiple of another lattice vector of $\sigma$.  From $\sigma^{\vee}$, one can define a semi-group $S=\sigma^{\vee}\cap M$, and the affine variety associated 
with $\sigma$ is 
\begin{equation}
X_\sigma=Spec(\sigma^{\vee}\cap M).
\end{equation}
We have the following further constraints on the cones:
\begin{itemize}
\item A strongly rational convex polyhedral cone (s.r.c.p.c) is a convex cone $\sigma$ such that $\sigma\cap (-\sigma)=0$. 
\item A simplicial s.r.c.p.c is a cone whose generator form a $R$ basis of $N$.  
\end{itemize}

We are interested in affine toric singularity which is defined by a cone $\sigma$. 
The singular locus of affine toric singularity is given by the sub-cone $\sigma_i$ whose generators do not form a Z-basis, and 
the singular locus corresponds to the orbit which is determined by $\sigma_i$.  For 3-fold 
toric singularity, the maximal dimension of singular locus is one dimensional which is then formed by a two dimensional sub-cone $\sigma_i$ in 
$\sigma$.  

A toric singularity is Q-Gorenstein if one can find a vector $m_\sigma$ in $M_Q$ such that $\langle m_\sigma, v_\rho\rangle=1$ for 
all the one dimensional generator $v_\rho$ of $\sigma$.  A toric singularity is Gorenstein if one can find a vector $H$ in the lattice $M$ such that
  $\langle H, v_\rho\rangle=1$.  Equivalently, one can choose a hyperplane in $N$ such that all the ray generators
lie on it.   For a 3-fold Q-Gorenstein singularity with 
index $r$, we take the hyperplane as $z=r$, and the ray generators form a two dimensional convex polygon $P$. Our 3-fold Q-Gorenstein 
toric singularity is then specified by an index $r$ and a two dimensional  convex polygon $P$.  The classification of 
3-fold canonical and terminal singularity are classified as follows:
\begin{itemize}
\item The cone $\sigma$ of 3-fold \textbf{canonical} singularity has the following property: there is no lattice points in $\sigma$ between the origin and the 
polygon $P$. In particular,  toric Gorenstein singularity is canonical. 

\item The cone $\sigma$ of 3-fold \textbf{terminal} singularity is characterized as follows: there is no lattice points in $\sigma$ between the origin and the 
polygon $P$, and furthermore there is no internal lattice points and boundary lattice points for $P$. 

\end{itemize}
 
We focus on isolated toric Gorenstein 3-fold singularity $X_P$ which is specified by a convex polygon $P$ with no lattice points on the boundary.   
The homology of the link $L_X$ is easy to compute. We ahve $\pi_1(L_X)=Z_3/\Gamma$, with $\Gamma$ the subgroup generated by the lattice vectors $v_\rho$; 
 $H_2(L_X,Z)$ has no torsion, and $b_2=n-3$ with $n$ the number of vertices of $P$. $H_3(L_X,Z)$ could have torsion though. Simply connected toric Sasakian manifold is  just $S^5 * k(S^2\times S^3)$. 

It was proven in \cite{futaki2009transverse} that the link associated with an isolated toric Gorenstein singularity admits Sasakian-Einstein metric. The corresponding Reeb vector field $\zeta$ can be determined using 
volume minimization \cite{martelli2008sasaki}.

\subsection{Hypersurface singularity} \label{hyper}
Let's consider a three dimensional isolated hypersurface singularity defined by the map $f:(C^4,0)\rightarrow (C,0)$. 
Isolated condition implies that equations $f={\partial f \over \partial z_i}=0$ has a unique 
solution at the origin. We also require that there is an effective $C^*$ action on the singularity: 
\begin{equation}
f(\lambda^{q_i}z_i)=\lambda f(z_i),~~q_i>0.
\label{quasi}
\end{equation}
Hypersurface singularity is Gorenstein, and the rational condition implies that 
\begin{equation}
\sum q_i -1>0.
\label{ration}
\end{equation}
All such rational hypersurface singularities have been classified in \cite{yau2005classification}. One define a five dimensional link $L_f$ of the singularity as the intersection with the standard five sphere:
\begin{equation}
L_f=X_f\cap S^5.
\end{equation}
Then $L_f$ has a positive Sasakian structure for $f$ satisfying condition \ref{quasi} and \ref{ration}. 

The topology of $L_f$ has been studied in mathematics literature, see \cite{boyer2005einstein}. We'd like to compute its homology group. 
One can define the Milnor fibration of the singularity. Let $\epsilon>0$ to be sufficiently small. The map $\phi: S_\epsilon^{5}/L_f\rightarrow S^1$ defined by 
\begin{equation}
\phi(z)={f(z)\over |f(z)|},
\end{equation}
is the projection map of a smooth fiber bundle with a smooth parallelizable fibre, and each fibre $F$ has the homotopy type of a bouquet of three spheres $S^3\vee\ldots \vee S^3$, and 
is homotopy equivalent to its closure $\bar{F}$ which is a compact manifold with boundary, where the common boundary $\partial \bar{F}$ is precisely $L_f$. Furthermore, $L_f$ is 
a smooth 1-connected manifold of dimension $5$, i.e., only $b_0, b_{2},b_{3},b_{5}$ are nonzero.  In particular, $L_f$ is simply connected. 

\textbf{Definition}: A $5$ dimensional manifold is called \textbf{homology} sphere if it has the same integral homology type as sphere: $H_0(M,Z)=H_{5}(M,Z)=1, H_i(M,Z)=0, i\neq 0, 5$. The homotopy group 
of homology sphere might be different from standard sphere. A \textbf{rational homology} sphere is defined similarly using rational coefficient instead of integer coefficient in homology group. 
A sphere is called \textbf{homotopy} sphere if it has the same homotopy type as standard sphere.

We would like to compute the Betti number and torsion part of homology group of the link $L_f$. 
The topology of the link is encoded by the following important exact sequence (here $F$ is the Milnor fibre, and $h_{*}$ is the monodromy group): 
\begin{equation}
0\rightarrow H_3(L_f,Z)\rightarrow H_3(F,Z)\overbrace{\rightarrow}^{1-h_{*}}H_3(F,Z)\rightarrow H_{2}(L_f,Z)\rightarrow 0.
\end{equation}
From this, we see that 
\begin{itemize}
\item $H_3(L_f,Z)=ker(1-h_{*})$ is a free abelian group.
\item $H_{2}(L_f, Z)=coker(1-h_{*})$ and in general it has torsion. 
\end{itemize}
The free part of the homology is encoded in the Alexander polynomial:
\begin{equation}
\Delta(t)=\text{det}(t I-h_{*}),
\end{equation}
and there is an effective  way to compute it.  
Let's start with an isolated rational hypersurface singularity with a $C^*$ action, we can choose the integral weights $(w_0,w_1,w_2, w_3)$ which has 
no common divisor, and the polynomial $f$ has weight $d$ instead of 1. Let's define the new set of 
rational numbers $({d\over w_0},{d\over w_2},\ldots,{d\over w_{3}})=({u_0\over v_0},\ldots, {u_n\over v_3})$, here $u_i$ and $v_i$ has no common divisor. We have 
\begin{equation}
b_{2}(L_f)=\sum(-1)^{4-s}{u_{i_1}\ldots u_{i_s}\over v_{i_1}\ldots v_{i_s} \text{lcm}(u_{i_1}\ldots,u_{i_s})}
\end{equation}
with the summation over $16$ subsets $(i_1,\ldots,i_s)$ of $(0,\ldots,3)$. 

\textbf{Example 1}: Consider the singularity $f=z_0^2+z_1^2+z_2^2+z_3^2$,  then we have $(w_0,w_1,w_2,w_3)=(1,1,1,1),~d=2$ and 
$u=(2,2,2,2),~v=(1,1,1,1)$. Then we have the subset $\emptyset, (0),(1), (2), (3)$, and $(01),(02), (03),(12),
(13),(23),(012),(013),(023),(123),(0123)$. Using above formula, we find 
\begin{equation}
b_2=1-4+2*6-4*4+8=1.
\end{equation}

\textbf{Example 2}: Consider the singularity $f=z_0^2+z_1^2+z_2^2+z_3^3$, then the weights are $(w_0,w_1,w_2,w_3)=(3,3,3,2),d=6$, so 
$u=(2,2,2,3)$ and $v=(1,1,1,1)$, we find 
\begin{equation}
b_2=1-4+3*3-14+4=0.
\end{equation}

The computation of torsion part of the homology is much more non-trivial, and here let's explain the formula of how to compute it. 
Given an index set $(i_1,\ldots, i_s)$, we will denote by $I$ all its $2^s$ subsets and by $J$ all its proper subsets. For each ordered subset $(i_1,\ldots, i_s)\subset (1,\ldots, 4)$ with 
$i_1<i_2<\ldots<i_s$, one defines inductively the set of $2^s$ positive integers, starting with $C_{0}=gcd(u_0,\ldots,u_n)$:
\begin{equation}
c_{i_1,\ldots,i_s}={gcd(u_0,\ldots,\hat{u_{i_1}},\ldots,\hat{u_{i_s}},\ldots,u_n)\over \prod_J c_{j_1\ldots j_t}}.
\end{equation}
 In addition, starting with $k_0=\epsilon_{4}$, one defines 
 \begin{equation}
 k_{i_1,\ldots,i_s}=\epsilon_{4-s+1}\sum_I(-1)^{s-t}{u_{j_1}\ldots u_{j_t}\over v_{j_1}\ldots v_{j_t} lcm(u_{j_1},\ldots, u_{j_t})}
 \end{equation}
where $\epsilon_{4-s+1}=0 (1)$,if $4-s+1$ is even (odd).

Now for any $1\leq j\leq r=[max(k_{i_1,\ldots,i_s})]$, we set 
\begin{equation}
d_j=\prod_{k_{i_1\ldots i_s \geq j}}c_{i_1,\ldots, i_s}
\end{equation}
The Orlik conjecture \cite{orlik1972homology} states that the torsion subgroup of $H_2$ is 
\begin{equation}
H_{2}(L_f,z)_{tor}=Z_{d_1}\bigoplus Z_{d_2} \ldots \bigoplus Z_{d_r}. 
\end{equation}

\textbf{Example}: The singularity is $f=z_0^2+z_1^3+z_2^5+z_2z_3^3$. The weights are $(15,10,6,8)$ and degree is 30. So we have $u=(2,3,5,15),v=(1,1,1,4)$.  We have 
\begin{align}
&c_{(0)}=1; c_{(1)}=1, c_{(2)}=1, c_{(3)}=1, c_{(4)}=1; c_{(12)}=5,c_{(13)}=3,c_{(14)}=1,c_{(23)}=1,c_{(24)}=1 \nonumber\\
&c_{(34)}=1,c_{(123)}=1,c_{(124)}=1,c_{(134)}=1,c_{(234)}=2,c_{(1234)}=1
\end{align}
For the number $k$s, we get
\begin{equation}
k=(0,0,0,0,-(3/4),0,0,0,0,0,0,0,0,0,2,0)
\end{equation}
So we have 
\begin{equation}
d_1=2,~d_2=2
\end{equation}
and $H_2(M,Z)_{tor}=Z_2\bigoplus Z_2$. 

For the singularity of the form $f=z_0^{a_0}+\ldots+z_3^{a_3}$, there is an easier way to compute the homology groups. To each polynomial with data $(a_0,\ldots, a_3)$, associate 
a graph $G(a)$ whose  vertices are labelled by $a_0,\ldots, a_3$. Two vertices $a_i$ and $a_j$ are connected if and only if $gcd(a_i,a_j)>1$. Let $G_{ev}(a) \in G(a)$ denote 
the connected component of $G(a)$ determined by the \textbf{even} integers. Then the following holds
\begin{itemize}
\item The link $L(a)$ is a rational homology sphere if and only if either $G(a)$ contains at least one isolated point, or $G_{ev}(a)$ has an odd number of vertices and for any distinct $a_i,a_j\in G_{ev}(a)$, $gcd(a_i,a_j)=2$.

\item The link $L(a)$ is a homology sphere if and only if either $G(a)$ contains at least two isolated points, or $G(a)$ contains one isolated point and $G_{ev}(a)$ has an odd number of 
vertices and for any distinct $a_i, a_j \in G_{ev}(a)$, $gcd(a_i,a_j)=2$. 

\end{itemize}

Since $L_f$ is simply connected, and we can identify it as one of Smale-Barden manifolds once we computed the homology groups of the link $L_f$, 

Having discussed the topological computation of the link associated with the hypersurface singularity. We would like to determine whether there will be SE metric. 
Let's now consider the implication of K stability on the existence of SE metric on the link $L$. It is difficult to construct  all test configurations, but it is 
always possible to generate one type test configuration whose consequence is just the unitarity bound from field theory perspective \cite{gauntlett2007obstructions, collins2012k}. 
Let $L(w,d)$ be a link of a weighted homogeneous hypersurface with weight vector $w=(w_0, w_1, w_2, w_3)$ ordered as $w_0\leq w_1\leq w_2 \leq w_3$, the generator of canonical bundle is 
\begin{equation}
\Omega={dx_0\wedge dx_1 \wedge dx_2 \wedge dx_3\over df},
\end{equation}
and it has weight $\sum w_i -d$. The normalization of the $C^*$ action is that $\Omega$ has charge two. So the $C^*$ charge of coordinate $ x_i$ 
is ${2w_i\over \sum w_i-d}$. The K stability implies that 
\begin{equation}
{2w_i\over \sum w_i-d} \geq {2\over 3}.
\label{uni}
\end{equation}
The above constraint is a necessary condition for the existence of SE metric on the link. It is also sufficient for the singularity $f=z_0^{a_0}+z_1^{a_1}+z_2^{a_2}+z_3^{a_3}$ if the exponents $a_i$ are pairwise coprime. 
We also have many other results by using K stability and sufficient conditions:
\begin{itemize}
\item Let $L(w,d)$ be a link of a weighted homogeneous hypersurface with weight vector $w=(w_0, w_1, w_2, w_3)$ ordered as $w_0\leq w_1\leq w_2 \leq w_3$. Let $Z_w$ denotes the corresponding weighted 
projective space. Furthermore let $I=\sum w_i -d$ denote the Fano index. Then (1): The 5-manifold $L(w,d)$ admits a Sasaki-Einstein manifold if $2Id<3 w_0 w_1$; 2) if the line $z_0=z_1=0$ does not lie in $Z_w$, and 
a weaker condition $2Id<3w_0w_2$ holds, then $L(w,d)$ admits a Sasaki-Einstein metric; (3) If the point $(0,0,0,1)$ does not lie in $Z_w$, and even weaker condition $2Id<3w_0 w_3$ holds, then $L(w,d)$ admits a Sasaki-Einstein metric.
Using this result, many SE manifold has been found in \cite{boyer2010sasaki}.
\item If $L(w,d)$  admits a $T^2$ action, we have following results \cite{collins2015sasaki}:
\begin{align}
& (1):~uv+z^p+w^q=0\rightarrow 2p>q~\text{and}~2q>p, \nonumber\\
& (2):~uv+z^p+zw^q=0\rightarrow  3(p-1)>(q+p-1)~\text{and}~2qp+1>p^2+q,\nonumber \\
& (3):~uv+z^pw+zw^q=0\rightarrow 3(p-1)^2(q-1)>(p+q-2)(pq-2p+1)~ \nonumber\\
&~~~~~~~~~~~~~~~~~~~~~~~~~~~\text{and}~3(q-1)^2(p-1)>(p+q-2)(pq-2q+1)
\label{hyperex}
\end{align}
The topology of them is respectively (1): $m(S^2\times S^3)$ with $m=gcd(p,q)-1$; 2) $m(S^2\times S^3)$ with $m=gcd(p-1,q)$; 3) $m(S^2\times S^3)$ with $m=gcd(p-1,q-1)+1$.
\item If $L(w,d)$  admits a $T^3$ action, there is no obstruction. The only example we know is conifold singularity $x_1^2+x_2^2+x_3^2+x_4^2=0$. 
\end{itemize}

\subsection{Complete intersection singularity}
The story can be easily generalized to an isolated singularity defined by  complete intersection. It is proven in \cite{reid1980canonical} that the rational Gorenstein 
complete intersection singularity has at most embedding dimension $5$. 
Consider an isolated complete intersection defined by two polynomials $f:(C^{5},0)\rightarrow (C^2,0)$. We require that complete intersection admits a $C^*$ action such that the weights are $(w_1, w_2,\ldots, w_{5})$ 
and the degrees are $(d_1,d_2)$. The rational condition implies that 
\begin{equation}
\sum_{i=1}^5 w_i -(d_1+d_2)>0.
\end{equation}
All these complete intersection singularities are classified in \cite{Chen:2016bzh}.  Consider a complete intersection singularity defined by two polynomials $f_1$ and $f_2$ with weights $(w_1,\ldots, w_5;d_1, d_2)$, one has a canonical three form
\begin{equation}
\Omega= {dx_1\wedge d x_2 \ldots \wedge d x_5\over df_1\wedge df_2}.
\end{equation}
The normalization is that it has charge 2 under the $C^*$ action, and the charge of each coordinate is constrained as follows:
\begin{equation}
{2 w_i \over \sum w_i -d_1-d_2} \geq {2\over 3},
\end{equation}
so that it might admit a SE metric, and the constraint comes from K stability.

\textbf{Example}: Consider the complete intersection singularity $(x_1^2+x_2^2+x_3^2+x_4^2+x_5^2,x_1^2+2x_2^2+3x_3^2+4x_4^2+5x_5^2)$. 
The corresponding link is a circle bundle over  Del Pezzo five ($dP_5$) surface.

\subsection{Quotient singularity}
Let's consider the quotient singularity of the form $C^3/G$ with $G$ a finite subgroup of of $SL(3)$. See appendix. \ref{app: quot} for the classification. 
The link is $S^5/G$. Only $G$ is abelian case, the singularity could be isolated, and these examples are toric so they admit SE metric.
The five dimensional link $S^5/G$ have been studied in many details, and they all admit SE metric. 

More generally, we can consider quotient singularity of above hypersurface and complete intersection singularity. To preserve the Gorenstein condition,
we require that the finite group action preserves the canonical three form. It would be interesting to study further these large class of examples.

\subsection{Minimal Gorenstein Log del Pezzo surfaces}
For a quasi-regular positive Sasakian manifold, we have a Seifert fibration structure $f: X\rightarrow (S, \Delta= \sum (1-{1\over m_i})D_i)$, here $S$ has cyclic quotient singularity and $-(K_S+\Delta)$ is 
ample. We would like to focus on one type of surface $S$ which is Gorenstein and $-K_S$ is ample. The Gorenstein condition implies that there are only cyclic du Val singularities (They are $A_n$ type surface singularities.).
The rank of these manifolds is defined as the rank of its Picard group.  There is no complete classification for such surfaces, but 
the rank one and rank two  relatively minimal Gorenstein log del Pezzo surfaces  are classified in \cite{miyanishi1988gorenstein, miyanishi1993gorenstein, ye2002gorenstein}, see table \ref{sconnected} and \ref{Nsm} for the full list of surfaces with cyclic singularities. 

Some deformation class of higher rank Gorenstein log Del Pezzo surfaces can be found as follows \cite{kollar2005einstein}. 
Let S be a Del Pezzo surface with Du Val singularities and $m_1\geq\ldots \geq m_k\geq 1$ integers. We denote by $B_{m_1,\ldots, m_k}S$ any surface obtained as follows: Pick any smooth elliptic curve $C\in |-K_S|$ and 
$p_1,\ldots, p_k$ distinct points on $C$. Then perform a blow up of type $m_i$ at $p_i$. All such surfaces form one deformation type. Furthermore, S and the deformation type determine the numbers $m_i$. Indeed, the number $m_i\geq 2$ can be 
read off from the singularities and the Picard number determines $k$. 
The canonical class of $B_{m_1,\ldots, m_k}S$ is nef and big iff $\sum m_i<K_S^2$. If this holds then $B_{m_1 \ldots m_k}S$ is a Del Pezzo surface for general choice of the points $p_i$. 
The above process does not change the \textbf{homology group} and \textbf{fundamental group}; and 
there are 93 deformation types of Del Pezzo surfaces with cyclic Du Val singularities satisfying $H_1(S)=0$ \footnote{For a singular surface $S$, the fundamental group is defined as the fundamental group of its smooth part $S^0$, and 
its homology group is the abelianization of the fundamental group.} . These are 
 \begin{itemize}
 \item $B_{m_1,\ldots,m_k}P(1,2,3)$ for $\sum m_i <6$ and $m_i\geq 2$.
 \item $B_{m_1\ldots, m_k}Q$ for $\sum m_i<8$ and $m_i\geq 2$.
 \item $B_{m_1,\ldots, m_k}P^2$ for $\sum m_i<9$ and $m_i\geq 2$.
 \item $B_{m_1,\ldots,m_k} P^1\times P^1$ for $\sum m_i<8$.
 \item $S_5,B_3S_5,B_4S_5$ and $B_1 P^2$.
 \end{itemize}
And all of them satisfy $\pi_1(S)=0$.  There are isomorphisms:
\begin{equation}
B_1P(1,2,3)=B_3 Q,~~B_1 Q=B_2 P^2,~~B_{m_1}P^2=B_m(P^1\times P^1),~~B_1S_5=B_5 P^2.
\end{equation}

Let's now study further the possible Sasakian structure built on above log Del Pezzo surfaces. 
Let's start with a log Del Pezzo surface $(S, \Delta=\sum(1-{1\over m})D)$, and we also include a $Q$ divisor $\Delta$. 
Recall that a Sasakian structure is determined by following data: a) Integers $0\leq b< m$ with $b$ coprime with $m$; b) A class of Weyl divisors $B$. 
The Chern class of Seifert bundle is 
\begin{equation}
c_1(X/S)=B +{b\over m} D.
\end{equation}
The smooth condition is that $m(x,\Delta)\cdot c_1(X/S)$ generates the local class group. Here $m(x,\Delta)$ is defined as 
$m(x,\Delta)=\{lcm(m_i) | x\in D_i\}$.  The corresponding Sasakian manifold has a Einstein metric if 
\begin{itemize}
\item The Chern class of Seifert bundle is a negative multiple of  Chern class of $(S,\Delta)$.
\item $(S,\Delta)$ has a orbifold Kahler-Einstein metric. 
\end{itemize}
The first condition is called pre-SE condition, and we also need to impose smoothness condition which significantly reduce the possible deformation class whose Sasakian manifold might have a SE metric. The proof goes as follows. 
For each of these rank one surface $T$, write $-K_T\sim d(T)H$ where $H\in Weil(T)$ is a positive generator, see table. \ref{sconnected}. We have 
\begin{equation}
d(P^1\times P^1)=2,~d(P^2)=3,~d(Q)=4,~d(S_5)=5,~d(P(1,2,3))=6.
\end{equation}
Next we perform some weighted blow ups to get $S=B_{m_1\ldots, m_k}T$. There are $k$ exceptional curves $E_1,\ldots, E_k$. Each $E_i$ 
passes through a unique singular point $p_i\in S$ and $E_i$ generates the local class group which is $Z/m_i$. Set $d(S)=gcd(m_1,\ldots,m_k, d(T))$. The divisor 
class group Weil(S) is freely generated by 
\begin{equation}
\pi^{*}H,~~E_1,\ldots,E_k,~~K_S=-d(T)\pi^{*}H+\sum m_i E_i.
\end{equation}
 In our case there is only one curve $D\sim -K_S$ and $S$ is smooth along D, and 
 \begin{equation}
 c_1(Y/X)=a \pi^{*}H+ \sum_i c_i E_i +{b\over m} D.
 \end{equation}
 Here $B=a \pi^{*}H+ \sum_i c_i E_i $ with the coefficients are all integers, and $b<m,~(b,m)=1$.
 Now pre-SE condition implies that $c_1(Y/S)=B+{b\over m}  D$ is 
a positive rational multiple of $-(K_S+(1-{1\over m})D)$. In our case $D\sim -K_S$, hence B itself is a rational multiple of $-K_S$, and 
\begin{equation}
B=-{ r\over d(S)}K_S=r({d(T)\over d(S)}\pi^{*}H-\sum{m_i\over d(S)}E_i).
\end{equation}
B generates the local class group 
at every singular point of T if and only $ m_i=d(S)=d(T)$ for all $m_i$. Thus 
$Y(B_{m_1,\ldots,m_k}T,B=a H+c_1 E_1+\ldots+c_k E_k, {b\over m} D)$ is smooth and pre-SE if and only if 
\begin{itemize}
\item $m_1=\ldots=m_k=d(S)$.
\item $aH+c_1 E_1+\ldots+c_k E_k=r(-{d(T)\over d(S)} H+E_1+\ldots+E_k)$ for some positive integer $r$ which is coprime with $d(S)$.
\item If T is singular, then $(r, d(T))=1$ and $d(S)=d(T)$. 
\end{itemize}
This cuts down the 93 deformation classes to the following 19 classes:
The pre-SE condition significantly reduces the above list and we have:
\begin{itemize}
\item $d(S)=1$:~~$B_1P^2$,~~$B_{11}P^2$,\ldots,$B_{11111111}P^2$
\item $d(S)=2$:~~$P^1\times P^1$, $B_2P_1\times P^1$, $B_{22}P^1\times P^1$, $B_{222}P^1\times P^1$.
\item $d(S)=3$: $P^2$,~$B_3P^2$,~$B_{33}P^2$.
\item $d(S)=4$:~$Q$,~$B_4Q$. 
\item $d(S)=5$:~$S_5$.
\item $d(S)=6$,~~$P(1,2,3)$.
\end{itemize}
And they are realized by the following hypersurface singularity \cite{kollar2009positive}:
\begin{itemize}
\item $B_2 P^1\times P^1$:~~$x^3+y^3+z^3+x t^m=0,~~(m,2)=1$.
\item $B_{22}P^1\times P^1$:~~$x^4+y^4+z^2+zt^m=0,~~(m,2)=1$.
\item $B_{222}P^1\times P^1$:~~$x^6+y^3+z^2+t^{3m},~~(m,2)=1$.
\item $B_3P^2$:~~$x^4+y^4+z^2+xt^m=0,~~(m,3)=1$.
\item $B_{33}P^2$:~~$x^6+y^3+z^2+zt^m=0,~~(m,3)=1$.
\item $Q$:~~$x^4+y^4+z^2+t^m=0,~~(m,2)=1$.
\item $B_4Q$:~~$x^6+y^3+z^2+yt^m=0,~~(m,2)=1$.
\item $S_5$:~~$x^6+y^3+z^2+zt^m=0,~~(m,5)=1$.
\item $P(1,2,3)$:~~$x^6+y^3+z^2+t^m=0,~~(m,6)=1$.
\end{itemize}
One can compute the homology group of these Sasakian manifold using the method studied in \ref{hyper} for hypersurface singularity, or the method listed in subsection one of this section. 
We could also use the K stability to check what kind of $m$ would give us a SE metric.  One simple consequence of  K stability is that $m$ can not be too large such that we can have a SE metric.

\begin{table}[h]
\begin{center}
  \begin{tabular}{ |c|c|c|c|c|}
    \hline
    degree & singularities & $\pi_1(S^0)$&Weil/Pic &univ.cover \\ \hline
        8 & $smooth$ & 0 & 1& $P^1\times P^1$ \\ \hline 
    9 & $smooth$ & 0 & 1& $P^2$ \\ \hline 
       8 & $A_1$ & 0 &Z/2& $Q$ \\ \hline 
              6 & $A_1+A_2$ & 0 &Z/6& $P(1,2,3)$\\ \hline 
       5 & $A_4$ & 0 &$Z/5$& $S_5$ \\ \hline 
  \end{tabular}
\end{center}
\caption{The list of  rank one and rank two simply connected Gorenstein del Pezzo surface. Here $S^0$ is the smooth part of $S$, degree is defined as $K_S^2$.  }
\label{sconnected}
\end{table}

\begin{table}[h]
\begin{center}
  \begin{tabular}{ |c|c|c|c|c|c|}
    \hline
    degree & singularities & $\pi_1(S^0)$&Weil/Pic &univ.cover &$\pi_1(S^0/D)$  \\ \hline
    1 & $A_8$ & Z/3 & Z/3& $B_{3111}P^2$ &Z/3\\ \hline 
       1 & $A_7+A_1$ & Z/4 &Z/4& $B_{2111}P^2$ &Z/4\\ \hline 
              1 & $A_5+A_2+A_1$ & Z/6 &Z/6& $B_{111}P^2$ &Z/6\\ \hline 
       1 & $4A_2$ & $(Z/3)^2$ &$(Z/3)^2$& $P^2$ &$G_{27}$\\ \hline 
       1 & $2 A_3+2A_1$ & $Z/2+Z/4$ &$Z/2+Z/4$& $P^1\times P^1$ &$G_{16}$\\ \hline 
       1 & $2A_4$ & Z/5 &Z/5& $B_{1111}P^2$ &Z/5\\ \hline 
       1 & $A_3+4A_1$ & $(Z/2)^2$ &$(Z/2)^3+(Z/4)$& $B_{22}P^1\times P^1$ &$Z/2+Z/4$\\ \hline 
       2 & $A_7$ & Z/2 &Z/4& $B_{41}P^2$ &Z/2\\ \hline 
       2 & $A_5+A_2$ & Z/3 &Z/6& $B_{11}Q$ &Z/3\\ \hline 
       2 & $2A_3+A_1$ & Z/4 &Z/2+Z/4& $P^1\times P^1$ &Z/2+Z/4\\ \hline 
       2 & $6A_1$ & $(Z/2)^2$ &$(Z/2)^4$& $P^1\times P^1$ &$G_8$\\ \hline 
       2 & $2A_3$ & $Z/2$ &Z/2+Z/4& $B_{22}P^1\times P^1$ &Z/4\\ \hline 
          3 & $A_5+A_1$ & $Z/2$ &$Z/6$& $B_3P^2$ &$Z/6$\\ \hline 
      3 & $3A_2$ & $Z/3$ &$(Z/3)^2$& $P^2$ &$(Z/3)^2$\\ \hline 
           4 & $A_3+2A_1$ & $Z/2$ &$Z/2+Z/4$& $Q$ &$(Z/2)^2$\\ \hline 
      4 & $4A_1$ & $Z/2$ &$(Z/2)^3$& $P^1\times P^1$ &$(Z/2)^2$\\ \hline 
  \end{tabular}
\end{center}
\caption{The list of  rank one and rank two  non-simply connected Gorenstein del Pezzo surface. Here $S^0$ is the smooth part of $S$, degree is defined as $K_S^2$.}
\label{Nsm}
\end{table}

On the other hand, the existence of orbifold Kahler-Einstein metric can be checked using algebraic methods \cite{kollar2005einstein}. 
We know that $P^1\times P^1$ and $P^2$ surface has Kahler-Einstein metric. For singular surface, 
the existence of KE metric on following degree one Gorenstein Del Pezzo surface has been established in \cite{kosta2009pezzo}: 
\begin{equation}
A_4, 2A_4,  A_4 +A_3, A_4 +2A_1, A_4 +A_1, A_3 +4A_1, A_3 +3A_1, 2A_3 +2A_1, A_3 +2A_1, A_3 +A_1, 2A_3, A_3,
\end{equation}
The above list denotes the singularity type of surface, see table. \ref{sconnected} and \ref{Nsm} for some explicit degree one examples.

\textbf{Remark}: The SE metric on link $L$ (or Ricci flat conic metric on the cone $X$) constructed using the orbifold Khaler-Einstein metric on $S$ is quasi-regular.
However, not all SE metric can come from this way. For example, it is possible that 
SE metric on $X$ is irregular and therefore we can not construct them using the orbifold Khaler-Einstein metric on the base.

\newpage

\section{AdS/CFT correspondence}
As reviewed in introduction, we have a correspondence between large N four dimensional $\mathcal{N}=1$ SCFT and type IIB string theory on following background: 
\begin{equation}
AdS_5\times L_X;
\end{equation} 
Here $L_X$ is a five manifold with a Sasaki-Einstein metric, and we also need to turn on $N$ unit of five form fluxes on $L_X$. $L_X$ is defined as the link of a three dimensional log terminal singularity $X$ which is K stable. The field theory is defined as 
the IR theory on $N$ D3 branes probing $X$. The main point of this section is to study properties of SCFT from various geometric 
objects associated with $L_X$.

Type IIB string theory has following bosonic massless fields: $(G_{\mu\nu}, B_{\mu\nu},\phi)$ which is frome gravity multiplet and Ramond-Ramond fields $C_0, C_2, C_4$, which are zero form, two form and four form fields \cite{green2012superstring}. Type IIB string theory also contains half-BPS non-perturbative objects $Dp$ brane with $p=-1,1,3,5,7,9$, and $NS5$ brane. 
In general, we actually have $(p,q)$ five branes, and $(p,q)$ strings \cite{polchinski1998string}. More interestingly, we have quarter BPS $(p,q)$ string junction or $(p,q)$ five brane junctions.  
One can learn many interesting physics about field theory by studying above objects in $AdS_5\times L_X$ background, and the geometry of $L_X$ plays a crucial role.

\subsection{Global symmetries} 
From AdS/CFT correspondence, the global symmetries are associated with the massless gauge fields in $AdS_5$ space. 
One can have massless gauge fields from isometry of SE metric \cite{green2012superstring}, 
and one can also get global symmetry
from massless mode of four form RR field $C_4$, and the number is given by Betti number $b_3$. Since the fundamental group of 
a SE manifold is finite, and so the first homology group is finite, we do not get continuous gauge fields by compactifying two form 
fields $B_2$ and $C_2$ using harmonic one form on SE manifold. 

\subsubsection{Isometry of SE manifold}
We know that the Reeb vector field $\zeta$  is a Killing vector field and generates an isometry group, and this corresponds to the $U(1)_R$ symmetry of the field theory. The precise relation is $U(1)_R={2\over 3} \zeta$, so the $\zeta$ charge is just the scaling dimension for a chiral operator.
 This isometry exists for any SE manifold  which matches the fact that $U(1)_R$ symmetry exists for any 4d $\mathcal{N}=1$ SCFT. In general, there could be more isometries. Here let's discuss some general facts.

Given a Riemannian manifold $(M,g)$ we let $Isom(M,g)$ denote the isometry group of $g$ which is the subgroup of diffeomorphism group of $M$ which 
leaves $g$ invariant. It is known that $Isom(M,g)$ is a finite dimensional Lie group and is a compact Lie group if $M$ is compact \cite{boyer2008sasakian}. 
For SE five manifold, we have the relation $1\leq dim(Isom(M,g))\leq 15$, and the upper bound is achieved for $S^5$ with standard Sasakian structure. For irregular SE manifold, the isometry group is at least two dimensional. 
We might also have finite isometry subgroup too.
 
It is not so easy to compute the full isometry group for SE manifold $L_X$. However, in many cases, 
global symmetries can be found from automorphism group $aut(X)$ of the singularity $X$, and the symmetry group are the maximal compact subgroup of $Aut(X)$. For  hypersurface singularity, most of those automorphism group is discrete (except the $C^*$ action which is identified with the $U(1)_R$ symmetry). The proof goes as follows: Let's start with a 
hypersurface singularity $f(z_0,\ldots,z_3)$, and consider the one parameter subgroup $\tau_t$ of $aut(f)$:
\begin{equation}
z_i^{'}=z_i+\sum_{j\geq 1} t^j g_{ij}(z_0,\ldots,z_n)
\end{equation}
 The condition of invariance is 
 \begin{equation}
 f(\tau_t(z_0),\ldots, \tau_t(z_3))=f(z_0,\ldots,z_3).
 \end{equation}
 Differentiating with respect to $z_i$ of above equation, and expanding in a Taylor series in $t$ and equating the coefficients of $t^j$ for each $j$ gives 
 \begin{equation}
 \sum g_{kj} {\partial f\over \partial z_k}=0.
 \end{equation}
Since the degree of $g_{kj}$ is $w_k$ and the degree of ${\partial f\over \partial z_k}$ is $d-w_k$, and not all ${\partial f\over \partial z_k}$
vanish or there is a cancellation for two different values of $k$, say $k$ and $l$. But the only way this happen is that $w_k=w_l$ and that both 
${\partial f\over \partial z_k}$ and ${\partial f\over \partial z_l}$ are linear in $z_k$ or $z_l$. This implies that $w_k=w_l={d\over 2}$. This 
implies that our singularity is $z_0^2+z_1^2+f(z_2,z_3)=0$, and only in this case there is more than one dimensional isometry group.

In the other direction, we know that toric SE manifold admits at least $U(1)^3$ isometry. 
For toric singularity, there is an elegant description of three abelian automorphism groups using the combinatorial tools, see \cite{cox2011toric}
for more details. For singularity with rank two automorphism group, there is also a combinatorial description, see \cite{altmann2012geometry}.

The analysis of discrete symmetry is more subtle, and a detailed analysis for many concrete models will be presented elsewhere. 

\subsubsection{Baryonic symmetry}
One can also get baryonic symmetry from RR four form field $C_4$, i.e. in the KK analysis, we consider field $C_{\mu mnp}$ with $\mu$ the index in AdS direction, 
and $m,n,p$ the direction in $SE_5$ direction. The number of such symmetries is equal to Betti number $b_3$. These symmetries 
are called Baryonic as the baryonic operator \footnote{If a dual gauge theory exists, one can construct them explicitly.} will be charged under these symmetries.

\subsection{Operator spectrum}
Using massless fields of type IIB string theory, we can perform a Klazu-Klein (KK) analysis and get various massive fields in five dimensional $AdS$ space, 
and the masses of these fields depend on the harmonic analysis on $L_X$. These KK fields would give us operators for the conformal field theory \cite{Witten:1998qj}.
We are interested in short multiplets of field theory. The Betti numbers $b_i$ of $L_X$ gives harmonic forms which give massless particles in $AdS_5$ space, and they will define some short multiplets. 
The other important fact is that a large class of chiral scalars are simply given by the coordinate ring of $X$, and this immediately gives us a lot of information about field theory.

\subsubsection{Harmonic analysis}
One can find scaling dimension of a boundary field theory operator from the mass of a particle in $AdS_5$ space. One 
can get the masses from Klazu-Klein analysis of type IIB string theory  on $SE_5$ manifold. The detailed
analysis is quite involved, see \cite{Kim:1985ez,ceresole2000spectrum,Eager:2012hx}. We are interested in short 
multiplets, and they have simple origins from cohomology groups of sheafs on singularity $X$, which is easier to compute. Here let's summarize the main result, see
table. \ref{spectrum}. First for a holomorphic function $f$ with charge $c$ under  $C^*$ action, one can get 
three chiral multiplets, and three semi-chiral multiplet \cite{Eager:2012hx}. They contribute to single trace index:
\begin{equation}
I_f=t^{2c}+t^{2c+6}.
\end{equation}
Now for a holomorphic one form with charge $r$ under $C^*$ action, there are associated three semi-chiral multiplet, and they 
contribute to single trace index \cite{Eager:2012hx}:
\begin{equation}
I_v=-t^{2r+6}.
\end{equation}

\begin{table}[h]
	\begin{center}
		\begin{tabular}{ |c|c|c|c|c|}
			\hline
			~ & Multiplet&Operator & $U(1)_R$&$\Delta$  \\ \hline
			$f$ &chiral &${\cal O}$ & ${2\over 3}c$ & $c$ \\ \hline 
			~ & ~&$\lambda_{\alpha}$ & ${2\over 3}c+1$ & $c+{3\over2}$ \\ \hline 
			~ & ~&$S$ &  ${2\over 3}c+2$ & $c+3$ \\ \hline 
			~ & Semi-chiral&${\cal O}_{\dot{\alpha}}$ &  ${2\over 3}c-1$ & $c+{3\over 2}$ \\ \hline 
			~ & ~&${\cal O}_{\alpha\dot{\alpha}}$ &  ${2\over 3}c$ & $c+3$ \\ \hline 					 
			~ & ~&$S_{\dot{\alpha}}$ &  ${2\over 3}c+1$ & $c+{9\over 2}$ \\ \hline 	
			$T^z$&Semi-chiral&	C & ${2\over 3}r$ & $r+2$			 					   \\ \hline
			~&~&	$C_{\alpha}$ & ${2\over 3}r+1$ & $r+{7\over 2}$ \\ \hline
			~&~&	$C^{'}$ & ${2\over 3}r+2$ & $r+5$ \\ \hline
		\end{tabular}
	\end{center}
	\caption{$f$ is a holomorphic function on $X$ with charge $c$ under $C^*$ action. $T^z$ is a holomorphic one form with charge $r$ under $C^*$ action.  }
	\label{spectrum}
\end{table}

The complex structure on $X$ satisfies the relation (here $\zeta$ is the Reeb vector field, and $\Psi=r{\partial \over\partial r}$):
\begin{equation}
I(\zeta)=\Psi,~~I(\Psi)=-\zeta.
\end{equation}
So $I(\Psi+i\zeta)=i(\Psi+i\zeta)$, i.e. $\Psi+i\zeta$ is anti-holomorphic vector field ($\bar{\partial}$ operator). 
A holomorphic function on the metric cone obeys the equation
\begin{equation}
(\Psi+i\zeta)f=0,
\label{hol}
\end{equation}
Let's now take $f=r^c\tilde{f}$ with $\tilde{f}$ has zero charge under $r{\partial \over \partial r}$, and $\zeta \tilde{f}= i c \tilde{f}$. $f$ now has charge c under the action of holomorphic vector field ${1\over 2}(\Psi-i\zeta)$ 
\begin{equation}
{1\over 2}(\Psi-i\zeta) f= c f.
\end{equation} 
So a holomorphic function $f$ on $X$ with charge $c$ would give a scalar function $\tilde{f}$ on SE manifold which has charge $ic$ under the Reeb vector field. One 
can further show that $\tilde{f}$ is actually an eigen-form on $L_X$, and the eigenvalue is $c(c+4)$.

To see how a holomorphic function $f$ on $X$ would give an eigenfunction of $L_X$, We will discuss more detail about the harmonic analysis on  Sasakian-Einstein manifold \cite{stromenger2010sasakian,Schmude:2013dua}. Let's start with a $2n+1$ Sasakian manifold $(M,\zeta, \eta, \Phi)$, and define the following operations:
\begin{align}
&L:~\Omega^p(M)\rightarrow \Omega^{p+2}(M),~~\alpha\rightarrow \omega \wedge \alpha  \nonumber\\
&\Lambda:~\Omega^p(M)\rightarrow \Omega^{p-2}(M),~\alpha \rightarrow \omega \lrcorner \alpha
\end{align}
Here $\omega={1\over 2} d\eta$, and $\lrcorner$ is the contraction operation.  
Now a differential form $\alpha \in \Omega^p(M)$ is called 
\begin{itemize}
\item \textbf{horizontal}~if~$\zeta  \lrcorner \alpha =0$,~~\textbf{vertical}~if~$\eta \wedge \alpha =0$, ~~\textbf{primitive}~if~$\Lambda \alpha=0$.
\end{itemize} 
For a $\alpha\in \Omega^p(M)$, we have:
\begin{itemize}
\item $h(\alpha)=\zeta \lrcorner (\eta \wedge \alpha) \in \Omega^p(M)(H)$ is called the horizontal part of $\alpha$. 
\item $v(\alpha)=\zeta\lrcorner \alpha \in \Omega^{p-1}(H)$  is called the vertical prat of $\alpha$.
\end{itemize}
Now we can decompose every form as follows
\begin{equation}
\alpha=h(\alpha)+\eta\wedge v(\alpha)
\end{equation}
and a form is denoted as a two component vector $\left(\begin{array}{c} \beta \\ \gamma \end{array}\right)$.
We have the standard differential operator $d$, and can define the operators $d^c:\Omega^p(M)\rightarrow \Omega^{p+1}(M)$ and $\delta^c:\Omega^p(M)\rightarrow \Omega^{p-1}(M)$ by
\begin{align}
& d^c:=\sum \phi e_i^* \wedge \nabla_{e_i}, \nonumber\\
&\delta^c:= -\sum\phi e_i \lrcorner \nabla_{e_i}.
\end{align}
where $e_i$ is the local basis of tangent bundle. The space of horizontal $p$ forms are denoted as $\Omega^p(M)$, and we can define the operator $d_H: \Omega^{p}(H)\rightarrow \Omega^{p+1}(H)$ as 
\begin{equation}
d_H:=h\circ d|\Omega^p(H),
\end{equation}
here $h\circ$ means we take the horizontal part.
Similarly we can define operator $d_H^c$ and $\delta_H^c$. The Laplacian acting on a $p$ form $\alpha= \left(\begin{array}{c} \beta \\ \gamma \end{array}\right)$ becomes 
\begin{equation}
\Delta = \left( \begin{array}{cc}\triangle_H-(L_\zeta)^2+4L\Lambda & -2 d_H^c \\ -2\delta_H^c& \triangle_H-(L_\zeta)^2+4L\Lambda  \end{array}\right).
\label{lap}
\end{equation}
We now define following differential operators using the complex structure $\Phi$:
\begin{equation}
\partial, \bar{\partial}:\Omega_C^p(H)\rightarrow \Omega_C^{p+1}(H),~~~\partial^*, \bar{\partial}^*:\Omega_C^p(H)\rightarrow \Omega_C^{p-1}(H).~~~
\end{equation}
They satisfy the following conditions
\begin{equation}
\partial^2=\bar{\partial}^2=(\partial^*)^2=(\bar{\partial}^*)^2=0,~[\partial, \bar{\partial}]=-2L L_\zeta,~[\partial^*, \bar{\partial}^*]=2\Lambda L_\zeta.
\end{equation}
So $\Omega^{k}(H)$ has a hodge decomposition $\Omega^{k}(H)=\bigoplus_{p+q=k} \Omega^{p,q}(H)$. We finally have 
\begin{align}
& d_H^c=-i(\partial-\bar{\partial}),~\delta_H^c=i(\partial^*-\bar{\partial}^*),  \nonumber\\
& \triangle_H=2\triangle_{\bar{\partial}}-i((2n+1)-2deg-1)L_\zeta,~~\triangle_H=\triangle_{\partial}+\triangle_{\bar{\partial}}.
\end{align}
Here $\triangle_{\bar{\partial}}=\bar{\partial}^*\bar{\partial}+\bar{\partial}\bar{\partial}^*$. In fact, we also have $\partial^*=iL\Lambda,~\bar{\partial}^*=-iL\Lambda$.
We are interested in  $p$ forms satisfying following conditions:
\begin{itemize} \label{harmonic}
\item primitive $\Lambda \alpha=0$.
\item horizontal $\zeta \lrcorner \alpha=0$. 
\item $\bar{\partial} \alpha = \bar{\partial}^* \alpha =0$, and in particular $\triangle_{\bar{\partial}} \alpha =0$. The condition $\bar{\partial}^* \alpha =0$ is actually redundant as the primitive condition already implies it as $\bar{\partial}^*=-iL\Lambda$ .
\end{itemize}
Combine with the relation $[\Lambda, \bar{\partial}]=-i\partial^*$ which holds on horizontal form, we see that $\partial^* \alpha=0$ and $\delta^c_H \alpha =0$ for primitive, horizontal and holomorphic forms.  
Now look at the Laplacian \ref{lap}, and its action on primitive and horizontal forms are
\begin{equation}
\Delta=2\Delta_{\bar{\partial}}-i((2n-2deg)L_\zeta-(L_\zeta)^2.
\end{equation}
Now we choose a primitive, horizontal and holomorphic $p$ form $\alpha$ such that $L_{\zeta}\alpha=ic \alpha$, 
we see that $\alpha$  is an eigenform: 
$\Delta \alpha= [2q(n-deg)+q^2 ]\alpha$, see \ref{lap}.  We have ( recall that we are interested in $n=2$) the eigenvalues of the forms: 
\begin{align}
&zero~form:~\lambda_0 = 4c+c^2 \nonumber\\
&one~form:~~\lambda_1=2c+c^2 \nonumber\\
&two~form:~~\lambda_2=c^2.
\end{align}
Moreover, for a zero form $\alpha$, we take $\tilde{\alpha}=r^c \alpha$ which is now holomorphic on the cone $X$ with charge $c$ under $C^*$ action, see \ref{hol}. So we 
see that a holomorphic function on the cone with charge $c$ would give an eigenfunction on $L_X$ with eigenvalue $H_0=c^2+4c$. This is one of 
most important relation between the cone $X$ and the geometry of $SE_5$ manifold. The scaling dimension of bottom component of vector multiplet I (see \cite{ceresole2000spectrum} page 18) is $E_0=\sqrt{H_0+4}-2=c$, so we get a chiral multiplet $\hat{B}_{{2\over 3}c,(0,0)}$!

On the other hand, for a primitive, horizontal, and holomorphic one-form $v$ with $\zeta$ charge $c$,  it can 
actually be written as $(df)^{\lambda}\Omega_{\lambda \mu}$ with $\Omega$ a holomorphic two form with $\zeta$ charge $3$ \cite{Eager:2012hx}. Scalar 
function $f$ is actually also an eigenfunction with $\zeta$ charge $c-3$, and its eigenvalues are $H_0=c^2+2c-3$. The scaling dimension of bottom component of vector multiplet I (see \cite{ceresole2000spectrum} page 18) is $E_0=\sqrt{H_0+4}-2=c-1$. So we get a semi-chiral multiplet  ${\cal C}_{{2\over 3}(c-3),(0,0)}$.

When $c=0$, those forms are called basic forms.
Each basic $p$ form also has a Hodge decomposition, and the dimension of harmonic solution is called basic Hodge number. 
Now, we have the following important theorem: in the positive Sasakian case, the $(p,0)$th and $(0,q)$th Hodge number vanishing for $p>0$ and $q>0$ \cite{goertsches2012rigidity}.
In the quasi-regular case, they are just the Hodge number of orbifold $(S,\Delta)$ \cite{goertsches2012rigidity}. So essentially the only nonzero basic Hodge
numbers are $(h^{0,0}, h^{1,1})$. $h^{0,0}=1$ and so the only non-trivial numbers are $h_{1,1}$ which determines the second and third Betti numbers of SE manifold. 

Let's now interpret those forms with $c\neq 0$ from base point of view, and we consider quasi-regular SE manifold, so we have an associated 
surface $(S, \Delta)$, and for simplicity, we assume that the branch divisor $\Delta$ is trivial here. 
We need to consider following cohomology groups:
\begin{equation}
\begin{split}
\oplus_n H^p(S, \Omega_S^q\otimes (-aK_S)^n),~~(p,q)=(0,0),(0,1),(1,1),(0,2).
\end{split}
\end{equation}
Here $a$ is a positive rational number that determines the Sasakian structure \footnote{Recall that the Sasakian structure constructed from a base surface
depends on a choice of Weyl divisor $B$, and Sasaki-Einstein condition implies that it is a rational positive multiple of $-K_S$, so $B=-a K_S$. Moreover, if we want a smooth Sasaki-Einstein manifold, $a$ should be chosen so that $-aK_S$ generates the class group at every point. }, we also use the vanishing theorem to constrain $p$ and $q$ \footnote{$H^p(S,\Omega^q\oplus L)=0$ for $p+q>2$ if $L$ is ample, in our case $L=-aK_S$ which is ample. Moreover, $H^{i}(S,K_S+D)=0, i>0$ if $D$ is ample, and in our case $D=-K_S$ is ample. We also have $H^i(S, {\cal O}_S)=0$ for our rational surface $S$. }. We now have cohomology groups:
\begin{equation}
H^0(S,  (-aK_S)^n).
\end{equation}
They contribute to three chiral multiplets and the lowest one is a chiral scalar with $U(1)_R$ charge $n$ and the contribution to single trace index is $t^{3n}$. 
The cohomology groups
\begin{equation}
\begin{split}
&H^0(S, \Omega_S^2\otimes (-aK_S)^n)=H^0(S, (-a n+1) K_S). \\
\end{split}
\end{equation}
contributes to three semi-chiral multiplets and the lowest one has $R$ charge $n-2$, so the contribution to single trace index is $t^{3n}$.
Finally, we have cohomology groups 
\begin{equation}
H^{0}(S, \Omega\otimes (-aK_S)^n )-H^{1}(S, \Omega\otimes (-aK_S)^n ),
\end{equation}
which gives three semi-chiral multiplets and the lowest one has $R$ charge $n-2$, so the contribution to single trace index is $-t^{3n}$. 
The single trace  index can then be computed using Riemann-Roch theorem. 

From the cone point view, we have the following cohomology groups:
\begin{equation}
H^0(X,{\cal O}_X)=\oplus_nH^0(S,(-aK_S)^n).
\end{equation}
So ${\cal O}_X$ is the structure sheaf of $X$, and $H^0(X,{\cal O}_X)$ is the coordinate ring of $X$. Similarly, the other nontrivial cohomology groups on the cone is $H^0(X,\Omega_X)$ with $\Omega_X$ the sheaf of one forms.

\textbf{Example}: Let's take $S=P^1\times P^1$, and $K_S=-2 l-2m$ with $l,m$ corresponding to two $P^1$ factors, and the nontrivial intersection numbers are 
$l\cdot l =2$ and $m\cdot m=2$, so $K_S^2=8$. We choose $a={1\over 2}$, and the cone $X$ is the 
conifold singularity $x^2+y^2+z^2+w^2=0$. The dimension of cohomology group (using Riemann-Roch theorem \footnote{For a divisor $D$, the Riemann-Roch theorem gives $\chi({\cal O}_S(D))={(D+K_S)\cdot D\over 2}+\chi({\cal O}_S)$, here $\chi({\cal O}_S(D))$ is the Euler number of the divisor $D$, and $\chi({\cal O}_S)$ is the Euler number of the manifold $S$. In our case $D={-n\over 2} K_S$, and $\chi({\cal O}_S)=1$.})is 
\begin{equation}
H^0(S, {\cal O}_S\otimes (-{1\over2}K_S)^n)=(n+1)^2.
\end{equation}
Each element of $H^0(S, {\cal O}_S\otimes (-{1\over2}K_S)^n)$ contributes to a holomorphic function with $R$ charge $n$ on the cone $X$.

\subsubsection{Hilbert series, central charge $a$ and $U(1)_R$ symmetry}
The $U(1)_R$ symmetry is identified with the Reeb vector field as follows: $U(1)_R={2\over 3}\zeta$. In the large N limit,
The central charge is inverse proportional to the volume of the SE manifold \cite{Martelli:2006yb}. 
The volume can be computed from the Hilbert series of the affine ring $X$. In practice, there are 
often many candidate $U(1)_R$ symmetry, and the way to determine it is to use volume 
minimization. If there is just one candidate $U(1)_R$ symmetry, we can use Hilbert series to compute the central charge $a$.

We would like to focus on Gorenstein ring, and so there is a generator $\Omega$ of the canonical bundle. 
$\Omega$ should have charge two under candidate $U(1)_R$ symmetry: 
\begin{equation}
\zeta(\Omega)=2. 
\end{equation}
This normalization ensures that $\zeta$ is actually the $U(1)_R$ symmetry.
We also require that the $\zeta$ action on the coordinates to be positive, and so we have a space of candidate $U(1)_R$ symmetries:
\begin{equation}
\zeta(x_i)>0,~~\zeta(\Omega)=2.
\end{equation}

In the large N limit, the central charge $a$ of field theory can be extracted from the Hilbert series of $X$ with respect to $C^*$ action. 
The Hilbert series of a graded ring is defined as follows:
\begin{equation}
H_\zeta(t)=\sum_{\zeta} t^{\alpha} dim H_\alpha.
\end{equation}
Here $H_\alpha$ is the subspaces of the ring with charge $\alpha$ under symmetry $\zeta$. For our case, this Hilbert series has the following expansion:
\begin{equation}
H(e^{-s})={a_0(\zeta)\over s^3}+{a_1(\zeta)\over s^2}+\ldots
\end{equation}
Due to the charge two condition on $\Omega$, we have $a_0=a_1$. $a_0$ is proportional to volume of Sasakian manifold with Reeb vector field ${3\over 2} \zeta$. The central charge of field theory is related to the coefficient $a_0$ as follows 
\begin{equation}
a(\zeta)={27\over 32}N^2{1\over a_0(\zeta)}.
\end{equation}
This is the leading order term of central charge $a$ in the large $N$ limit. 

The Hilbert series for a complete intersection singularity is rather easy to compute. Let's start with a hypersurface singularity $f$, and  assume the weights are $(w_1,\ldots, w_4;d)$, then the Hilbert series is simply
\begin{equation}
H(t)={(1-t^d)\over (1-t^{w_1})(1-t^{w_2})(1-t^{w_3})(1-t^{w_4})}.
\end{equation}
The canonical differential is given by 
\begin{equation}
\Omega={dx\wedge dy \wedge dz \wedge d w\over df},
\end{equation}
and it has charge $\sum w_i-d$. Normalize the $C^*$ action so that $\Omega$ has charge two, so the normalized weights are ${2\over \sum w_i-d}(w_1,\ldots, w_4;d)$, and we find
\begin{equation}
a_0=\frac{d (-d+w_1+w_2+w_3+w_4)^3}{8  w_1w_2 w_3 w_4}.
\label{centralhyper}
\end{equation}

The Hilbert series of  a complete intersection singularity with weights $(w_1,w_2,w_3,w_4,w_5;d_1,d_2)$ is
\begin{equation}
H(t)={(1-t^{d_1})(1-t^{d_2})\over (1-t^{w_1})(1-t^{w_2})(1-t^{w_3})(1-t^{w_4})(1-t^{w_5})}.
\end{equation}
and the normalized weights are ${2\over \sum w_i-\sum d_i}(w_1,w_2,w_3,w_4,w_5;d_1,d_2)$ by requiring $\Omega= {dx_1\wedge \ldots \wedge dx_5\over df_1\wedge df_2}$ to have charge two. We find that
\begin{equation}
a_0=\frac{d_1 d_2 (-d_1-d_2+w_1+w_2+w_3+w_4+w_5)^3}{8  w_1 w_2 w_3 w_4 w_5}.
\end{equation}

\textbf{Example}: Consider the conifold example. The affine ring is simply given by the ideal $f=x^2+y^2+z^2+w^2$, and the weights are $(1,1,1,1;2)$, and we find (use \ref{centralhyper}):
\begin{equation}
a={27\over 64}N^2.
\end{equation}

Since Hilbert series plays a crucial role in determining the $U(1)_R$ symmetry and central charge, we would like to make some further comments on the general ring considered in this paper. For our SE manifold, we associate 
an affine ring $X$ which is further rational Gorenstein. For a  general ring $R$ with finite number of generators $y_i$ with charge $e_i$ ($i=1,\ldots,n$), the Hilbert series has the following general structure \cite{stanley1978hilbert}:
\begin{equation}
F(R,t)={P(R,t)\over \prod_i (1-t^{e_i})}.
\end{equation}
Here $P(R,t)$ is a polynomial with integer coefficients. There is a unique number $d$ such that $lim_{t\rightarrow 1} (1-t)^d F(R,t)$ is nonzero, and this $d$ is the dimension of the ring. Here 
we consider ring with dimension $d=3$, and the Hilbert series has an expansion
\begin{equation}
F(R,e^{-s})={a_0\over s^3}+{a_1\over s^2}+\ldots
\end{equation}
here $a_0= {e\over \prod_i e_i}$ and $e$ is the leading order coefficient of expansion of $P(R,t)=e(1-t)^{n-3}+\ldots$. 
Now we would like to restrict to the Gorenstein ring, the Hilbert series satisfy the following condition:
\begin{equation}
F(R,{1\over t})=(-1)^dt^{\rho}F(R,t).
\end{equation}
Here $d$ is the dimension and $\rho$ is an integer. For a Gorenstein ring, we have a regular sequence with length $d$. 
Let $\theta_1,\ldots, \theta_d$ to be a homogeneous regular sequence of $R$, and we have $deg(\theta_i)=f_i$, and let $S=R/(\theta_1,\ldots, \theta_d)$, then 
$S$ is a zero dimensional Gorenstein algebra: $S=S_0+\ldots+S_s$. The Hilbert series has the following nice form \cite{stanley1978hilbert}:
\begin{equation}
F(R,t)={(h_0+h_1 t+\ldots +h_s t^s) \over \prod_{i=1}^d (1-t^{f_i})}.
\end{equation}
where $h_i=dim(S_i)$.  There is a perfect pairing $h_i=h_{s-i}$, and furthermore $\rho=\sum f_i-s$. So it seems that the regular sequence plays 
an important role, and it would be interesting to understand their meanings in field theory.

We have discussed the consequence of Gorenstein condition on Hilbert series. The rational condition implies that $a_0=a_1>0$.

\subsubsection{Mesonic and baryonic operators}
Let's consider $N$ D3 brane probing three dimensional canonical singularity $X$. The moduli space of field theory on D3 branes
is equal to the symmetric product of X: $M_{vac}=X^N/S_N$. One part of chiral ring of the field theory is conjectured to be 
the coordinate ring of $M_{vac}$, and one may call them \textbf{mesonic operator}. These chiral operators might be separated into two parts: single trace and multiple trace operators. The name comes from the gauge theory where the chiral operators can be formed from trace of fundamental fields in the Lagrangian (This is the way to produce gauge invariant objects.): the single (multiple) trace operators are formed from a single (multiple) trace. 

In the large N limit, the space of single trace scalar chiral operators parameterizing the moduli space have a very nice description:
\begin{equation}
\textbf{The space of single trace chiral operators are described by the affine ring of $X$.}
\end{equation}
Because of the $C^*$ action, this affine ring is a graded ring $R=R_0\oplus R_{n_1}\ldots$, here $R_0=C$. Each subspace $R_{n_i}$ gives rise to a subspace of 
scalar chiral operator with charge $n_i$. The $C^*$ action is proportional to $U(1)_R$ charge, so one can find the scaling dimension of these chiral scalar operators from the grading of $X$. 
The multiple trace chiral operators are formed by simply multiplying 
the single trace operators. The chiral ring relation is completely captured by the ring $X$. 

\textbf{Example}: Let's consider conifold example which is defined by an ideal $f=x_1^2+x_2^2+x_3^2+x_4^2$, and each coordinate $x_i$ has R charge $1$, so we have the graded ring $R=R_0\oplus R_1\oplus R_2 \ldots$, i.e. $R_1$ has four elements generated by $x_i$; for $R_2$ we can form $9$ elements $x_ix_j$, but we have one quadratic relation, so there are 8 independent elements in $R_2$.

Let's now consider non-perturbative chiral operators in the large N limit. These operators are derived from wrapped D branes. 
We have \textbf{baryonic chiral operators} \cite{Berenstein:2002ke,Intriligator:2003wr,Herzog:2003dj} derived by wrapping $D3$ branes on three cycles.
The BPS condition is that the three cycles combined with $r$ direction is a holomorphic surface in the cone $X$ \cite{Beasley:2002xv,Mikhailov:2000ya}. 

For a quasi-regular SE manifold, one can find some of those surfaces from the divisors of the bases. 
Consider a quasi-regular Sasakian-Einstein manifold, and then we have a Seifert fibration $f: X\rightarrow (S,\Delta)$, 
here $S$ has cyclic quotient singularity, and $\Delta=\sum (1-{1\over m_i})D_i$ is a $Q$ divisor.  
 One can get baryons by wrapping D3 branes on effective divisors of $S$ (combining with fibre direction, we get a three cycle), and the lift of three cycle to the cone $X$ would be holomorphic, and we expect them to give BPS baryons. Now let's choose a basis of Weyl divisors $D_i$ on $S$, and consider a divisor $ D=\sum_{i=1}^n a_i D_i$, and the scaling dimension of these baryons are \cite{Herzog:2003dj}:
\begin{equation}
\Delta=-3N{K\cdot D \over K\cdot K}.
\end{equation}
Here $K=K_S+\Delta$ is the canonical divisor of the orbifold $(S,\Delta)$. If $D$ is effective ($a_i\geq 0$), the above number is positive as $-K$ is ample, which we might 
call baryons. On the other hand, if $a_i\leq 0$, we should get anti-baryon. The detailed counting of these baryons needs further study. 

If the first homology is nontrivial, one could also get particles by wrapping F1 string and D1 string on one cycles inside SE manifold. These baryons seem to 
have nilpotent ring relation as the first homology group is finite for SE manifold.

\subsection{Deformation}
\subsubsection{Exact marginal deformations}
It is interesting to study various deformations of the field theory through the deformation of $L_X$. Let's first consider exact marginal deformations:
\begin{enumerate}
\item We have  type IIB dilaton-axion $\tau$ in ten dimension, and this always give an exact marginal deformation of field theory. Type IIB string theory has a $SL(2,Z)$
duality acting on $\tau$, so all of these SCFTs also have a $SL(2,Z)$ duality symmetry.
\item We also get exact marginal deformations by compactifying 
two forms $B_2$ and $C_2$ if $b_2$ is nonzero. There are a total of 
$b_2$ (second Betti number) exact marginal deformations.
\item Moduli space of Sasaki-Einstein metric \cite{green2012superstring}. It is also interesting to study S duality on this moduli space.
\end{enumerate}
While the number of first two class of deformations are relatively easy to compute, the third class of exact marginal deformation is more difficult. For 
hypersurface singularity, however, it is easy to count the number, i.e. 
they are given by weight one deformations. 

\textbf{Example}: Let's consider the singularity $f=x^2+y^2+z^4+w^4$, which 
admits SE metric through section 4, see \ref{hyperex}. There is a weight one 
deformation of $f$, i.e. $F=f+\tau x^2y^2$, so the corresponding SCFT has one extra exact marginal deformation from the moduli of SE metric.

Notice however that not all exact marginal deformations are captured by geometric properties of $SE_5$ manifold; Some of exact marginal deformations 
are captured by turning on other fluxes besides the usual five form flux, i.e. so-called $\beta$ deformation \cite{Lunin:2005jy}.  Moreover, one can find other multiple 
trace exact marginal deformations too, and one can find these numbers by computing superconformal index.

\subsubsection{Deformation and resolution of singularity} \label{deform}

Given a singularity, one has two ways to make a singularity non-singular: deformation or resolution. For an isolated rational Gorenstein singularity, we have two special kinds of deformation called mini-versal deformation
and crepant resolution, and they should correspond to supersymmetric deformations.

The mini-versal deformation for a hypersurface singularity can be easily described. Let's start with a hypersurface singularity $f:(C^4,0)\rightarrow (C,0)$ which is quasi-homogeneous:
\begin{equation}
f(\lambda^{q_i} z_i)=\lambda f(z_i).
\end{equation}
The canonical differential has the form:
\begin{equation}
\Omega = {dz_1\wedge dz_2\wedge dz_3\wedge dz_4 \over df}.
\end{equation}
The normalization condition implies that $\Omega$ has charge $2$, and we have the normalization constant $\delta$ such that 
\begin{equation}
\delta (\sum q_i-1)=2 \rightarrow \delta={2\over \sum q_i-1}.
\end{equation}
The mini-versal deformation of the singularity is just:
\begin{equation}
F(z,\lambda)=f(z)+\sum_{\alpha=1}^\mu \lambda_\alpha \phi_{\alpha}.
\end{equation}

What is the field theory interpretation of these deformations? 
Here $\phi_\alpha = \prod z_i^{n_i}$ is the monomial basis of the Jacobi algebra $J_f$ of $f$ \footnote{$J_f$ is defined as the space ${C[z_1,z_2,z_3,z_4]\over \{{\partial f\over \partial z_1},\ldots,{\partial f\over \partial z_4}\}}$.}.  Each $\phi_{\alpha}$ has $U(1)_R$ charge $R(\phi_\alpha)={2\sum n_i q_i \over \sum q_i-1}$, and the parameter $\lambda_\alpha$ has $U(1)_R$ charge $R(\lambda_\alpha)={2(1-\sum n_i q_i) \over \sum q_i-1}$. The deformation can be classified by the scaling dimension of $\lambda_\alpha$:
\begin{itemize}
	\item Relevant deformation: $[\lambda_\alpha] >0$ or $\sum n_\alpha q_\alpha<1$.
	\item Marginal deformation: $[\lambda_\alpha]=0$ or $\sum n_\alpha q_\alpha=1$.
	\item Irrelevant deformation: $[\lambda_\alpha]<0$ or $\sum n_\alpha q_\alpha>1$.
\end{itemize}
For hypersurface singularity, generically there is no global symmetry besides the $R$ symmetry, and the marginal deformations are actually exact marginal \cite{Green:2010da}. So we conclude that the weight one deformation of the singularity gives the exact marginal deformation. 

We can turn on relevant deformation and ask what is the IR SCFT. For the hypersurface singularity, one can actually determine the IR SCFT by using 
similar trick used in \cite{Xie:2015rpa}. The idea is by simply taking a scaling limit of the deformed polynomial and get a new quasi-homogeneous singularity from which one can read IR SCFT. 

\textbf{Example}: Consider the singularity $f=x_1^2+x_2^2+x_3^2$, and $x_4$ is free. The field theory is four dimensional $\mathcal{N}=2$ affine $A_1$ quiver. 
The addition of any monomials to $f$ is flat and is a deformation. A particular simple deformation is 
$f^{'}=x_1^2+x_2^2+x_3^2+x_4^{2k}$. The deformation in the field theory side can be described by adding a superpotential deformation $\text{Tr}(\Phi_1^k)+\text{Tr}(\Phi_2^k)$ \cite{Cachazo:2001sg} to the affine $A_1$ quiver corresponding to singularity $f=x_1^2+x_2^2+x_3^2$.

For resolution of singularity, we have a special kind of resolution called crepant resolution. Given a morphism $f: Y\rightarrow X$, and it is a  crepant resolution if  
\begin{equation}
K_Y=f^{*} K_X.
\end{equation}
It is proven that for 3-fold canonical singularity $X$, one can always have a partial crepant resolution such that $Y$ is Q-factorial and has only terminal singularity. Such resolutions are not unique, but the number of crepant divisor is the same! For toric singularity, one can describe the resolution in a very explicit way \cite{cox2011toric}, for crepant resolution 
of other hypersurface singularity, see \cite{Xie:2017pfl}.

\textbf{Example}: For hypersurface singularity $f=z_1^2+z_2^2+z_3^2+z_4^2$, there is a crepant resolution such that the exceptional set is just a curve \cite{reid1980canonical}. On the other hand, hypersurface singularity $f=z_1^2+z_2^2+z_3^2+z_4^3$ is already Q-factorial and is a terminal singularity, so it does not admit a crepant resolution.   

From the field theory point of view,  
it seems that the corresponding field theory for the resolved geometry might be described by fractional branes \cite{Klebanov:2000hb}. Namely, if a quiver gauge theory is known for the original singularity $X$, then the field theory for resolved geometry might be described by changing the rank of the quiver nodes.  
See \cite{Franco:2005zu,Pinansky:2005ex,Butti:2006hc} for more  discussions on this issue. The dynamics of these deformed field theory is very interesting and we plan to study them in the future. 

\subsection{Extended objects}
We have mainly discussed how to find the information of local operators of field theory from geometry of SE manifold , and now we would like to discuss the extended object of field theory.
Let's first summarize some topological properties of SE manifold.
The fundamental group of a manifold with positive Ricci curvature is finite, and so the first and fourth integral homology group of SE manifold is finite. 
For simply connected SE manifold, the torsion group of $H_2(L_X,Z)$ is quite constrained. One can get extended objects by wrapping branes on 
these non-trivial homology cycles:

\textbf{Strings}: One can get strings in $AdS_5$ space by: a): unwrapped F1 and D1 string; b) wrapping $D3$ brane one two cycles; c): wrapping $D5$, $NS5$ brane on four cycles. These objects are related to line operators of field theory. 

\textbf{Membranes}: One can get membranes in $AdS_5$ space by: a): wrapping $D3$ brane on one cycle; b): wrapping $D5$, $NS5$ brane on three cycles. These 
objects are related to surface defects of field theory.

\textbf{Domain walls}: One can get domain walls in $AdS_5$ space by: a): unwrapped D3 branes; b): wrapping $D5$, $NS5$ branes on two cycles; c): wrapping D7 brane on four cycle. These objects are related to domain wall of field theory.

More generally, one can get extended BPS objects in $AdS_5$ space \cite{Yamaguchi:2003ay} by wrapping various $p$ branes on calibrated cycles in $SE_5$. The calibrated cycle in $SE_5$ is defined using its extension to the cone: i.e. An odd dimensional cycle in $SE_5$ is 
holomorphic if its extension to the cone $X$ is holomorphic; and a two dimensional cycle in $SE_5$ is special Lagrangian if its three dimensional 
extension to $X$ is Special Lagrangian (they are called Legendrian submanifold). So we could have calibrated one cycle, two cycle and three cycle in 
SE manifold. 

If one can find Legendrian submanifold (which is two cycle in $SE_5$), one can wrap a $D5$ brane to get a BPS domain wall for field theory. 
We actually already studied baryons by wrapping $D3$ branes one three cycles 
which is holomorphic in the cone $X$, see more discussions in \cite{Mikhailov:2000ya,Beasley:2002xv}. If we use $D5$ brane on these three cycles, we should get BPS surface defects of field theory.

In general, one could also turn on gauge fields on branes and it is a very interesting question to classify all stable branes on a SE manifold. We leave these quesitons for the future study.

\section{A conjecture about reducing checking K stability to finite cases}
The space of SE manifolds is already quite large due to the results using various sufficient condition methods, see the summary in section 4. The necessary and sufficient condition 
for the existence is the K stability. While the sufficient condition is technically hard and not optimal, one can actually prove the existence of SE metric for many interesting cases. 
K stability is technically easier to implement for a given test configuration, but potentially infinite number of test configuration makes it very difficult to get new SE metrics. The known 
success of using K stability comes from reducing the number of non-trivial test configuration to possibly finite cases, and this is possible mainly for the case with many symmetries, for example, 
for the toric case, one do not need to consider non-trivial test configuration.  To enlarge the space of SE metric by using the method of K stability, one need to have some new insights to 
reduce the number  of test configuration to possibly finite numbers, and the purpose of this section is to propose a conjecture on such reduction based on intuition from field theory.

Let's recall that the definition of K stability of a ring $X$ involves a test configuration and the computation of Futaki invariant. The test configuration is 
described by a one parameter family of rings such that: a: for $t\neq 0$, $X_t$ are all isomorphic; b):
the central fibre $X_0$ is different from $X_t$ for non-trivial test configuration, and $X_0$ has one more dimensional automorphism group if $X_0\neq X_t$. The 
Futaki invariant is computed by using Hilbert series: 
\begin{equation}
H(s)={a_0\over s^3}+{a_1\over s^2}+\ldots
\end{equation}
In particular, a test configuration with central fibre $X_0$ destablizes $X$ if one can get a lower $a_0$ from $X_0$ (Notice that the range of parameter in symmetry space is restricted to positive half of the new symmetry $\eta$. ). One could reduce the space of test configurations by using symmetries (if $X$ is toric, then no destablizing nontrivial test configuration.), and requiring $X_0$ to be normal (i.e. the singularity locus of $X_0$ is at most one dimensional.). However, in general, it seems that there are infinite number of test configurations that we need to check.

\begin{figure}[h]
	\centering
	\includegraphics{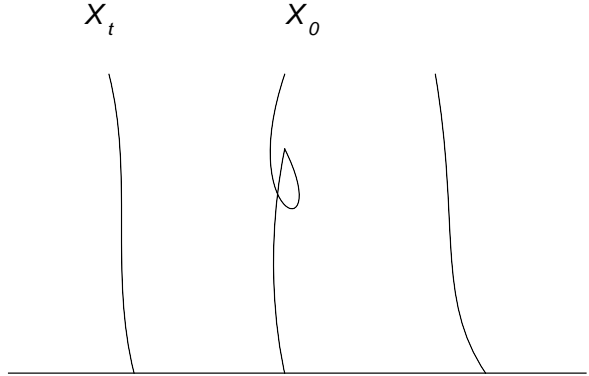}
	\caption{A test configuration is a family of rings parameterized by $t$. $X_t$ for $t\neq 0$ are isomorphic, while $X_0$ is different from $X_t$ for non-trivial test configuration.}
	\label{test}
\end{figure}

Let's now interpret the ingredients of K stability in terms of field theory terms. As we discussed in last section, $X$ essentially determines the chiral ring of field theory, and this fact is also true even if $C$ is not K stable.
Since K stable singularity gives a SE metric, which in turn defines a four dimensional $\mathcal{N}=1$ SCFT, so from physical point of view, K stability is equivalent to check that whether $X$ is the chiral ring of a SCFT!

Now the test configuration can be thought of as a one parameter deformation 
of a theory associated with $X_0$, so finding a test configuration is equivalent to find a $UV$ theory and a one parameter deformation such that the chiral ring of deformed theory is described by $X_t$. 
Let's illustrate this point in more detail. 
Consider the flat family parameterized by $t$ and the central fibre is $X_0$, $X_0$ is constrained to be also rational Gorenstein so that we still have a local field theory 
from $D3$ branes probing the singularity $X_0$. From $X_0$ point of view, $X_t$ is just one parameter deformation of $X_0$. From physical point of view, 
$X_t$ is derived as a deformation of field theory associated with $X_0$ \footnote{One should be careful that sometimes ${\cal O}$ is not an element of chiral ring of $X_0$ as it might involve the fractional power of certain 
chiral operators, but the physical picture is still correct.}:
\begin{equation}
X_t=X_0+t\int d^4x d^2\theta {\cal O} +c.c.
\end{equation} 
Here ${\cal O}$ is a chiral operator of theory associated with $X_0$ \footnote{Here we use superpotential to describe the deformation, and it is possible that the deformation is described by giving an expectation value to certain chiral operators, however, in that case, the deformation is necessarily relevant deformation.}. We have following two scenarios:
\begin{enumerate}
\item if ${\cal O}$ is relevant, then the theory flows to a new SCFT, and the chiral ring of the IR SCFT \textbf{might} be described by $X_t$, but 
it might be described by a ring $X_\infty$ which is different from $X_t$. We are definitely sure that $X_0$ is not the chiral ring of IR SCFT associated with $X_t$.
\item if ${\cal O}$ is marginal or irrelevant,  then $X_t$ would flow back to theory associated with $X_0$, and $X_0$ would have larger central charge $a$ as 
it is the UV theory of $X_t$. This also means that $X_t$ can not be the chiral ring of the IR SCFT. 
\end{enumerate}
So from field theory point of view, checking $K$ stability is essentially checking whether $X_t$ is the chiral ring of the IR SCFT for theory defined using $X_t$. 
K stability involves finding all possible RG flow from a $UV$ theory $X_0$ and a one parameter deformation $t$ such that the chiral ring of  deformed theory is given by $X_t$. The deformation necessarily breaks some symmetries of the $UV$ theory as ${\cal O}$ would be charged under some symmetries of $X_0$. 
$X_0$ destablize $X_t$ if the interaction is marginal or irrelevant. In principle, the above procedure can only tell us the necessary condition for $X_t$ to be the chiral ring of the IR SCFT, but in this context, it is also the sufficient condition!

Now here comes one crucial point, from $X_t$ point of view, ${\cal O}$ would have $R$ charge two! So it appears that one only need to study the test configuration involving (perhaps only single trace) chiral operators of theory $X_t$ with $R$ charge two! 

Let's now implement above observations to hypersurface case. Let's start with an isolated quasi-homogeneous hypersurface singularity $f:(C^4,0)\rightarrow (C,0)$. 
As we discussed in section \ref{deform}, the deformation parameter before each weight one monomial in $f$ has $R$ charge zero, and it should be associated with an operator with $R$ charge two. So if our above physical observations are true, one only need to consider test configuration involving these weight one monomials.   
 
In hypersurface case, there is a nice way of finding such test configuration.  
Let's recall how to find an isolated hypersurface singularity with $C^*$ action. First, we have to include one of the monomials in the set $(z_0^{a_0},z_1z_0^{a_0},z_2z_0^{a_0},z_3z_0^{a_0})$ in $f(z_0,z_1,z_2,z_3)$, otherwise $f$ would 
have a non-isolated singularity in $z_0$. Similarly, we need to include one of monomials in each of the following four sets 
\begin{equation}
\begin{split}
&I=(z_0^{a_0},z_1z_0^{a_0},z_2z_0^{a_0},z_3z_0^{a_0}),~~~~~~~II=(z_1^{a_1}z_0,z_1^{a_1},z_2z_1^{a_1},z_3z_1^{a_1}), \\
&III=(z_0z_2^{a_2},z_1z_2^{a_2},z_2^{a_2},z_3z_2^{a_2}),~~~~IV=(z_3^{a_3}z_0,z_3^{a_3}z_1,z_2z_3^{a_3},z_3^{a_3}) 
\end{split}
\end{equation}
We call the monomials within above four sets as extremal monomials. Once we pick four monomials from above sets, one can figure out the weights $w_i$ of coordinates $z_i$, and these weights cut a hyperplane $H$ in $R^4$. One might also add other monomials if there are other \textbf{positive integral} points on $H$. Generically the polynomial has just one dimensional $C^*$ action $\zeta$.  Now let's consider 
a test configuration which is generated by a $C^*$ action $\eta$, and the central fibre $X_0$ should have a two dimensional symmetry group, which implies 
that the singularity of $X_0$ is not isolated (generically). Let's now assume that we would like to make $z_0$ axis to be a singular locus of $X_0$, we only need to find a test configuration so that one of  term in set $I$ of $f$ has highest weight under the symmetry $\eta$. Let's choose $\eta$ which acts on the coordinates as 
$z_i\rightarrow t^{b_i}z_i$, and it is always possible to choose $b_i$ such that $z_j z_0^{a_0}$ term has weight one and other monomials having weight zero,  so this monomial dropped out in defining equation of $X_0$. In practice, there might be more than one terms from set $I$ which is compatible with the weights, and we treat them separately as the action $\eta$ would be different. Now we would like to make the following conjecture:
\begin{prop}
	For isolated hypersurface singularity $f(z_0,z_1,z_2,z_3)$, one only need to consider test configurations which eliminate one of extremal monomials in $f$. 	
\end{prop}	
The above conjecture is true for $f=z_0^{a_0}+z_1^{a_1}+z_2^{a_2}+z_3^{a_3}$ if the exponents $a_i$ are pairwise coprime, since the monomials appear in 
the equation are the only allowed monomial, and the test configuration which eliminates one of the extremal monomial will give the constraint \ref{uni}, which 
is also sufficient for the existence of SE metric \cite{boyer2008sasakian}.
It is also true for the known examples listed in section 4. It would be interesting to further study above conjecture for general case.

\section{Conclusion}
We initiated a study of AdS/CFT correspondence between  four dimensional $\mathcal{N}=1$ SCFT and newly discovered five dimensional Sasaki-Einstein manifold $L_X$.
One could get exact results of field theory from the geometric 
properties of SE manifold. Given that not much is known about the dynamics of general $\mathcal{N}=1$ SCFT, we believe this class of theories would provide us 
a very good framework to learn four dimensional $\mathcal{N}=1$ SCFT and general properties of AdS/CFT correspondence.

Although the metric is not known explicitly,  many field theory properties can be learned by using algebraic geometry tools. On the one hand, we have a three dimensional affine ring $X$ which determines a log terminal singularity. For a quasi-regular $SE$ manifold, one can associate a log del Pezzo surfaces $(S, \Delta)$. Many questions about 
$SE$ manifold can be reduced to the study of $X$ and $(S, \Delta)$, which is often much easier to deal with as there are many algebriac geometry tools available. 
One can greatly expand the list of $SE$ manifolds by using $K$ stability, and now we have a much larger space of four dimensional $\mathcal{N}=1$ SCFTs. 
The really nice feature of this class of theories is that many exact questions of field theory can be reliably computed from geometry, which will give us many insights about four dimensional $\mathcal{N}=1$ SCFT. 

Previous studies are mostly focusing on 
toric singularity (see the review \cite{Kennaway:2007tq,Yamazaki:2008bt}) where many features are not generic, for example, in general case, the SE metric would have moduli, the deformation of singularity is more interesting,
and homology group is more rich, etc. More studies are needed to understand the field theory implications of these new geometric features. 

We have made some conjectures on testing K stability of complete intersection singularity. It would be interesting to 
prove whether above conjecture is correct or not. In particular, it would be nice to find out all K stable singularities
from table. \ref{table:isolatedsingularitiesALEfib}. 

In this paper, We mainly use SE manifold to define our field theory and learn their properties through 
. In second part of this series, we will mainly use field theory tools to 
study AdS/CFT correspondence. 

\section*{Acknowledgements}
We would like to thank Tristan Collins and Yifan Wang for helpful discussions.
The work of S.T Yau is supported by  NSF grant  DMS-1159412, NSF grant PHY-
0937443, and NSF grant DMS-0804454.  
The work of DX is supported by Yau Mathematical Science Center at Tsinghua University. D.X would like to thank Caltech for hospitality
at the final stage of the completion of this paper.

\appendix{}

\section{Three dimensional quasi-homogeneous hypersurface singularity} \label{app: hyper}
The three dimensional quasi-homogeneous hypersurface singularity has been classified in \cite{yau2005classification}, and we reproduce it here:
\begin{center}
  \begin{tabular}{|l|c|c| }
    \hline
    Type & $f(z_0,z_1,z_2,z_3)$ & $\sum q_i$  \\ \hline
    I & $z_0^a+z_1^b+z_2^c+z_3^d$& ${1\over a}+ {1\over b}+{1\over c}+{1\over d}$\\ \hline
       II & $z_0^a+z_1^b+z_2^c+z_2z_3^d$& ${1\over a}+ {1\over b}+{1\over c}+{c-1\over cd}$\\ \hline
       III & $z_0^a+z_1^b+z_2^cz_3+z_2z_3^d$& ${1\over a}+ {1\over b}+{d-1\over cd-1}+{c-1\over cd-1}$\\ \hline
       IV & $z_0^a+z_0z_1^b+z_2^c+z_2z_3^d$& ${1\over a}+ {a-1\over a b}+{1\over c}+{c-1\over cd}$\\ \hline
       V & $z_0^az_1+z_0z_1^b+z_2^c+z_2z_3^d$& ${b-1\over ab-1}+ {a-1\over a b-1}+{1\over c}+{c-1\over cd}$\\ \hline
       VI & $z_0^az_1+z_0z_1^b+z_2^cz_3+z_2z_3^d$& ${b-1\over ab-1}+ {a-1\over a b-1}+{d-1\over c}+{cd-1\over cd}$\\ \hline
       VII & $z_0^a+z_1^b+z_1z_2^c+z_2z_3^d$& ${1\over a}+ {1\over b}+{b-1\over b c}+{b(c-1)+1\over bcd}$\\ \hline
       VIII & $z_0^a+z_1^b+z_1z_2^c+z_1z_3^d+z_2^pz_3^q$,  & ${1\over a}+ {1\over b}+{b-1\over b c}+{b-1\over bd}$\\ 
 ~ &  ${p(b-1)\over bc}+{q(b-1)\over bd}=1$& ~\\  \hline
       IX & $z_0^a+z_1^bz_3+z_2^cz_3+z_1z_3^d+z_1^pz_2^q$,  & ${1\over a}+ {d-1\over bd-1}+{b(d-1)\over c(bd-1)}+{b-1\over bd-1}$\\ 
 ~ &  ${p(d-1)\over bd-1}+{qb(d-1)\over c(bd-1)}=1$& ~\\  \hline
        X & $z_0^a+z_1^bz_2+z_2^cz_3+z_1z_3^d$& ${1\over a}+ {d(c-1)+1\over bcd+1}+{b(d-1)+1\over bcd+1}+{c(b-1)+1\over bcd+1}$\\ \hline
                XI & $z_0^a+z_0z_1^b+z_1z_2^c+z_2z_3^d$& ${1\over a}+ {a-1\over ab}+{a(b-1)+1\over abc}+{ab(c-1)+(a-1)\over abcd}$\\ \hline

                XII & $z_0^a+z_0z_1^b+z_0z_2^c+z_1z_3^d+z_1^pz_2^q$& ${1\over a}+ {a-1\over ab}+{a-1\over ac}+{a(b-1)+1\over abd}$\\ 
                ~&${p(a-1)\over ab}+{q(a-1)\over ac}=1$&~ \\ \hline
  XIII & $z_0^a+z_0z_1^b+z_1z_2^c+z_1z_3^d+z_2^pz_3^q$& ${1\over a}+ {a-1\over ab}+{a-1\over ac}+{a(b-1)+1\over abd}$\\ 
                ~&${p(a(b-1)+1)\over abc}+{q(a(b-1)+1)\over abd}=1$&~ \\ \hline
                  XIV & $z_0^a+z_0z_1^b+z_0z_2^c+z_0z_3^d+z_1^pz_2^q+z_2^rz_3^s$& ${1\over a}+ {a-1\over ab}+{a-1\over ac}+{a-1\over ad}$\\ 
                ~&${p(a-1)\over ab}+{q(a-1)\over ac}=1={r(a-1)\over ac}+{s(a-1)\over ad}$&~ \\ \hline
                XV&$z_0^az_1+z_0z_1^b+z_0z_2^c+z_2z_3^d+z_1^pz_2^q$& ${b-1\over ab-1}+{a-1\over ab-1}+{b(a-1)\over c(ab-1)}+{c(ab-1)-b(a-1)\over cd(ab-1)}$ \\ 
                ~&${p(a-1)\over ab-1}+{qb(a-1)\over c(ab-1)}=1$&~ \\ \hline
                 XVI&$z_0^az_1+z_0z_1^b+z_0z_2^c+z_0z_3^d+z_1^pz_2^q+z_2^rz_3^s$& ${b-1\over ab-1}+{a-1\over ab-1}+{b(a-1)\over c(ab-1)}+{b(a-1)\over d(ab-1)}$ \\ 
                ~&${p(a-1)\over ab-1}+{qb(a-1)\over c(ab-1)}=1={r(a-1)\over ac}+{s(a-1)\over ad}$&~ \\ \hline
                    XVII&$z_0^az_1+z_0z_1^b+z_1z_2^c+z_0z_3^d+z_1^pz_2^q+z_0^rz_2^s$& ${b-1\over ab-1}+{a-1\over ab-1}+{a(b-1)\over c(ab-1)}+{b(a-1)\over d(ab-1)}$ \\ 
                ~&${p(a-1)\over ab-1}+{qb(a-1)\over d(ab-1)}=1={r(b-1)\over ab-1}+{sa(b-1)\over c(ab-1)}$&~ \\ \hline
                XVIII&$z_0^az_2+z_0z_1^b+z_1z_2^c+z_1z_3^d+z_2^pz_3^q$&${b(c-1)+1\over abc+1}+{c(a-1)+1\over abc+1}+{a(b-1)+1\over c(abc+1)}+{c(a(b-1)+1)\over d(abc+1)}$ \\
                ~&${p(a(b-1)+1)\over abc+1}+{qc[a(b-1)+1]\over d(abc+1)}=1$&~ \\ \hline
                XIX &$z_0^a+z_0z_1^b+z_2^cz_1+z_2z_3^d$ &${[b(d(c-1)+1)-1\over abcd-1}+{[d(c(a-1)+1)-1\over abcd-1}$\\
                ~&~&$+{[a(b(d-1)+1)-1\over abcd-1}+{[c(a(b-1)+1)-1\over abcd-1}$ \\
                
    \hline
  \end{tabular}
\end{center}

\section{Three dimensional Gorenstein quotient singularity}\label{app: quot}
Three dimensional Gorenstein quotient singularity has been classified in \cite{yau1993gorenstein}, and we reproduce it here:
\begin{itemize}
\item (A) Diagonal abelian groups. These can be described using toric geometry.
\item (B) Group isomorphic to transitive finite subgroups of $GL(2, C)$.
\item (C) Group generated by (A) and T.
\item (D) Group generated by (C) and Q.
\item  (E) Group of order 108 generated by S, T, V.
\item (F) Group of order 216 generated by (E) and $P = UVU^{-1}$.
\item (G) Hessian group of order 648 generated by (E) and U.
\item (H) Simple group of order 60 isomorphic to alternating group $A_5$.
\item (I) Simple group of order 168 isomorphic to permutation group generated
by $(1234567)$, $(142)(356)$, $(12)(35)$.
\item (J) Group of order 180 generated by (H) and F .
\item (K) Group of order 504 generated by (I) and F .
\item (L) Group G of order 1080 its quotient $G/F$ isomorphic to alternating group $A_6$ , where $F = \{I,W,W^2\}$ is the center of $SL(3, C)$,
$I=$ identity. 
\end{itemize}
The matrices used above are
\begin{equation}
S=\left(\begin{array}{ccc}
1&0&0 \\
0&\omega&0\\
0&0&\omega^2 \\
\end{array}\right),~
T=\left(\begin{array}{ccc}
0&1&0 \\
0&0&1\\
1&0&0 \\
\end{array}\right),~
V={1\over \sqrt{-3}}\left(\begin{array}{ccc}
1&1&1 \\
1&\omega&\omega^2\\
1&\omega^2&\omega \\
\end{array}\right),~
\end{equation}

\begin{equation}
U=\left(\begin{array}{ccc}
\epsilon&0&0 \\
0&\epsilon&0\\
0&0&\epsilon \omega \\
\end{array}\right),~
P={1\over \sqrt{-3}}\left(\begin{array}{ccc}
1&1&\omega^2 \\
1&\omega&\omega\\
\omega&1&\omega \\
\end{array}\right),~
Q={1\over \sqrt{-3}}\left(\begin{array}{ccc}
a&0&0 \\
0&0&b\\
0&c&0 \\
\end{array}\right),~
\end{equation}
with $abc=-1,\omega=e^{2\pi i \over 3}, \epsilon^2=\omega^2$. The A and B type matrices are
\begin{equation}
A=\left(\begin{array}{ccc}
\alpha&0&0 \\
0&\beta&0\\
0&0&\gamma\\
\end{array}\right),~
B=\left(\begin{array}{ccc}
\alpha&0&0 \\
0&a&b\\
0&c&d \\
\end{array}\right),~
W=\left(\begin{array}{ccc}
\omega&0&0 \\
0&\omega&0\\
0&0&\omega \\
\end{array}\right),~
\end{equation}

We have $\alpha\beta\gamma =1$ and $ad-bc={1\over \alpha}$.

\bibliographystyle{utphys}

\bibliography{PLforRS}

\end{document}